\pdfoutput=1

\documentclass[iop]{emulateapj}
\usepackage{natbib}
\usepackage{graphicx}
\usepackage{epstopdf}
\usepackage{footnote}

\shorttitle{Variability Power Spectra of PKS\,0735+178}
\shortauthors{Goyal et al.}

\begin{document}

\title{Multi-wavelength variability study of the classical BL Lac object PKS\,0735+178\\ on timescales ranging from decades to minutes}

\author{Arti Goyal\altaffilmark{1},
{\L}ukasz Stawarz\altaffilmark{1},
Micha{\l} Ostrowski$^1$,
Valeri Larionov$^2$,
Gopal-Krishna$^3$, 
Paul~J.~Wiita$^4$ ,
Santosh Joshi$^5$, 
Marian Soida$^1$,
and Iv\'an Agudo$^6$.
}

\altaffiltext{1}{Astronomical Observatory of Jagiellonian University, ul.\ Orla 171, 30-244 Krak\'ow, Poland}
\altaffiltext{2}{Astronomical Institute of St.\ Petersburg State University, Petrodvorets 198504, Russia}
\altaffiltext{3}{Centre for Excellence in Basic Sciences (CEBS), University of Mumbai campus (Kalina), Mumbai 400098, India}
\altaffiltext{4}{Department of Physics, The College of New Jersey, 2000 Pennington Rd., Ewing, NJ 08628-0718, USA}
\altaffiltext{5}{Aryabhatta Research Institute of Observational Sciences (ARIES), Manora Peak, Nainital 263002, India}
\altaffiltext{6}{Instituto de Astrof\'{\i}sica de Andaluc\'{\i}a (CSIC), Apartado 3004, E--18080 Granada, Spain}

\email{email: {\tt arti@oa.uj.edu.pl}}

\begin{abstract}

We present the results of our power spectral analysis for the BL Lac object PKS\,0735+178 utilizing the {\it Fermi}-LAT survey at high-energy $\gamma$-rays, several ground-based optical telescopes, and single-dish radio telescopes operating at GHz frequencies. The novelty of our approach is that, by combining long-term and densely sampled intra-night light curves in the optical regime, we were able to construct for the first time the optical power spectrum of the blazar for a time domain extending from 23 years down to minutes. Our analysis reveals that: (i) the optical variability is consistent with a pure red noise, for which the power spectral density can well be approximated by a single power-law throughout the entire time domain probed; (ii) the slope of power spectral density at high-energy $\gamma$-rays ($\sim $1), is significantly flatter than that found at radio and optical frequencies ($\sim $2) within the corresponding time variability range; (iii) for the derived power spectra we did not detect any low-frequency flattening, nor do we see any evidence for cut-offs at the highest frequencies down to the noise floor levels due to measurement uncertainties. We interpret our findings in terms of a model where the blazar variability is generated by the underlying single {\it stochastic} process (at radio and optical frequencies),  or a linear superposition of such processes (in the $\gamma$-ray regime). Along with the detailed PSD analysis, we also present the results of our extended (1998--2015) intra-night optical monitoring program and newly acquired optical photo-polarimetric data for the source.

\end{abstract}


\section{Introduction} 
\label{sec:intro}

The blazar class of active galactic nuclei (AGN) includes BL Lacertae objects (BL Lacs) and high polarization quasars \citep[HPQs, which are a subset of flat spectrum radio quasars, FSRQs, characterized by high fractional optical polarization degree ${\rm PD_{opt}}> 3\%$; see][]{1995PASP..107..803Urry}. For these, the total radiative energy output is dominated by the broad-band, non-thermal emission produced in relativistic jets.  These jets are launched by supermassive black holes surrounded by magnetized accretion disks, that are located at the centers of (usually) massive elliptical galaxies, and when pointing close to our line of sight the emission is substantially Doppler boosted \citep{Begelman84,DeYoung02,Meier12}. The radio-to-optical/X-ray segment of the blazar emission continuum is due to the synchrotron radiation of ultrarelativistic leptons (hereafter ``electrons'', for simplicity), while the high-frequency X-ray-to-$\gamma$-ray segment is most widely believed to be due to the inverse-Comptonization of various circumnuclear photon fields (produced both internally and externally to the outflow) by the jet electrons.

Blazars display a strong variability from radio to $\gamma$-rays on timescales ranging from decades down to hours, or even minutes. The observed flux changes are often classified broadly into the three major types, namely a `long-term variability' (LTV, with the corresponding timescales of decades-to-months), a `short-term variability' (STV; weeks-to-days), and an intra-night/day variability \citep[INV/IDV; timescales less than a day; see, e.g.,][]{1995ARA&A..33..163W,Ulrich97,2014A&ARv..22...73F}. Rapid -- hour- and minute-long -- blazar flares are especially pronounced at higher energies, particularly in the $\gamma$-ray regime, with intensity changes of up to even a few magnitudes \citep{2007ApJ...664L..71A,2011ApJ...730L...8Aleksic,2016ApJ...824L..20A,Foschini11,Saito13,Rani13}. The origin of such dramatic behavior, and in particular its relation to the similarly rapid but smaller-amplitude `microvariability' observed at lower wavelengths, (including INV in the optical range), is still being widely debated. 

Several competing scenarios have been proposed to explain the rapid variability of blazar sources. Some include {\it extrinsic causes}, such as gravitational lensing \citep[where the observed flux changes are attributed to a lensing of light rays by a foreground compact object; e.g.,][]{Schneider87,1991Natur.349..766G,Webb00} or, in the radio domain, interstellar scintillation \citep[see][for a review]{Melrose94}. Others involve {\it an intrinsic origin}, including some that are purely geometrical in nature, such as `the lighthouse effect', where the precession of a jet results in the differential forward beaming of the emission \citep{1992A&A...255...59C,1992A&A...259..109GK}. A set of promising intrinsic origin models, which may be favored because of polarimetric measurements revealing the accompanying changes in the fractional polarization and polarization angle, involves various plasma instabilities leading to the formation of shocks and turbulence in the outflow, which then heat and accelerate the jet particles \citep[e.g.,][and references therein]{2012MNRAS.420..604N,2012MNRAS.423.1707S,2014ApJ...780...87M, 2015ApJ...809..171S}.  Other intrinsic models invoke the annihilation of magnetic field lines of opposite polarity, transferring the energy from the field to particles at the reconnection sites and during the subsequent evolution of the reconnected magnetic field \citep[e.g.,][]{2009MNRAS.395L..29G, 2015MNRAS.450..183S,2016ApJ...818L...9G}.

At optical frequencies, the LTV of non-blazar AGN, including radio-quiet quasars (RQQs), radio-intermediate quasars (RIQs), lobe-dominated radio-loud quasars (LDQs), and low-polarization flat spectrum radio quasars (LPQs), is characterized by much smoother, yet comparable in amplitude, intensity changes to those observed in blazar sources \citep[e.g.,][]{2005MNRAS.356..607S, 2011ApJ...743L..12M, 2013ApJ...766...16E}. This finding is surprising, keeping in mind that the bulk of the optical emission of these non-blazar type active galaxies is understood to originate in accretion disks, rather than relativistic jets \citep[e.g.,][]{2006ASPC..350..183W}. On the other hand, a clear dichotomy exists on intra-night timescales, where the blazar class shows a considerably higher optical INV in both amplitude, $\psi$ (see equation \ref{eq:psi} below), and duty cycle, DC (equation \ref{eq:DC}), than non-blazar objects \citep[e.g.,][]{2003ApJ...586L..25GK, 2005MNRAS.356..607S, 2005A&A...440..855Gupta, 2009MNRAS.397.1893V, 2010MNRAS.401.2622G}. More recently, based on a systematic study using 262 intra-night light curves, each monitored for a duration of $\geq$ 4 hr, it was shown that the optical INV duty cycle is DC\,$\sim 40\%$ for the blazar class, while it is  $\sim 5\%$, $\sim 11\%$, $\sim 3\%$,  and $\sim 10\%$ for the RQQs, RIQs, LDQs, and LPQs, respectively, at least whenever clear INV amplitudes of $\psi \geq 3\%$ are considered \citep{2013MNRAS.435.1300AG}.  

Even though the physical processes giving rise to the flaring emission of blazars remain debatable, considerable progress has been made in characterizing the statistical properties of blazar variability at different wavelengths, and in different time domains. In particular, it has been demonstrated repeatedly that the power spectral density (PSD) of blazar light curves is, in general, of a power-law form \citep{1985ApJ...296...46S,2001ApJ...560..659K,2003A&A...397..565P,2003A&A...402..929B,2007ApJ...664L..71A,2007A&A...467..465C,2008ApJ...689...79C,2012ApJ...749..191C,2010ApJ...722..520A,2011AJ....141...49C,2011A&A...531A.123K,2013ApJ...766...16E,2013ApJ...773..177N, 2014ApJ...786..143S,2014ApJ...785...60R,2014ApJ...785...76P,2014MNRAS.445..437M,2015A&A...576A.126A, 2015ApJ...798...27I, 2015MNRAS.451.4328K}. A physical process with such a variability power spectrum, denoted hereafter as $P(\nu_k)=A \, \nu_k^{-\beta}$, where $\nu_k$ is the temporal frequency (corresponding to the timescale $1/\nu_k$), $A$ is the normalization constant, and $\beta$ is the spectral slope, is called white noise when $\beta = 0$, flicker (pink) noise when $\beta = 1$, and Brownian (red) noise  when $\beta = 2$ \citep{1978ComAp...7..103P}. The PSD integrated over some variability frequency range is then a measure of the variance of the underlying signal in the time series within the corresponding range of variability timescales. Breaks in the slope or in the normalization of a PSD may appear, signaling characteristic/critical variability timescales in the system. In the case of blazars, various segments of radio, optical, X-ray, and $\gamma$-ray power spectra within the variability time domains from years to days (and in some instances, even sub-hour timescales), are characterized by spectral slopes $1 \leq \beta < 3$, meaning that the variability amplitude increases with increasing variability timescale. Rarely, however, have blazar PSDs  been analyzed in a systematic way at different wavelengths across the electromagnetic spectrum and over a truly broad range of temporal frequencies. It is important to note that colored noise-type power spectra are expected to flatten on longer variability timescales (to preserve the total finite variance), and to cut-off at frequencies corresponding to the shortest variability timescale in a system. The detection of such cutoffs in blazar periodograms would be of a primary importance for constraining the physics of blazar jets; however, such detections may be hampered by the finite duration of available monitoring blazar data, on the one hand, and statistical fluctuations resulting from the measurement errors, on the other hand.

In this work, we present our analysis and interpretation of the multi-wavelength (radio, optical, and high-energy $\gamma$-ray), and particularly long-span (decades to minutes) variability power spectrum of the BL Lac object PKS\,0735$+$178. This source is singled out from the blazar class by its persistently weak intra-night optical variability: using 17 nights data spanning over 11 years of optical monitoring of this blazar, \citet{2009MNRAS.399.1622AG} estimated the INV DC as $\sim 0\%$ for $\psi > 3\%$. Note that while such a low duty cycle is not unusual for non-blazar AGN (see above), it is surprising for highly polarized blazars such as PKS\,0735$+$178. In this study, we present the results of our extended intra-night monitoring programme of the source, now consisting a total of 25 nights and spanning over 18\,yr of observations (1998--2015).

PKS\,0735$+$178 \citep[J2000.0 R.A.\,$\rm=07^{h}38^{m}07\fs39$, Dec.\,$\rm=+17\degr42\arcmin19\farcs99$; redshift $z = 0.45\pm0.06$;][]{2012A&A...547A...1Nilsson} is an otherwise typical example of a low-frequency-peaked BL Lac object detected in the GeV photon energy range \citep{2010ApJ...715..429Abdo}. It is highly polarized in the optical band \citep[${\rm PD_{opt}} > 3\%$;][]{1996AJ....112.1877G, 2011ApJS..194...19W}, and exhibits a flat-spectrum radio core with a superluminal pc-scale radio jet, both characteristic of  blazars in general \citep{1992ApJ...398..454Wills, 2016AJ....152...12L}.  Despite its pronounced optical and radio variability on yearly timescales \citep{1988AJ.....95..374W, 2007A&A...467..465C}, the source is relatively quiet in X-rays \citep{1988ApJ...330..776M}, and the existing rather sparse X-ray monitoring data preclude a meaningful power spectral analysis. Within the high-energy $\gamma$-ray regime covered by {\it Fermi}-LAT, the blazar is detected at a high significance level only with weekly or longer temporal binning \citep{2003ApJ...597..615N,2010ApJ...722..520A}. Upper limits for the PKS\,0735$+$178 emission in the very high-energy $\gamma$-ray domain (photon energies $>100$\,GeV) have been recently provided by the VERITAS Collaboration \citep{2016AJ....151..142A}.

\begin{table*}[th!]
\caption{Observational summary of the INV data and data analysis}
\label{tab:result}
\tiny
\centering
\begin{tabular}{ccccccccccccccc}\\\hline
Date of obs. & Tel.     & Dur.     &  $N_p$   & $\Delta m_{t-s1}$  & $\Delta m_{t-s2}$ & $\Delta m_{s1-s2}$  & $\sigma_{s1-s2}$ & SD$_{s1-s2}$    & $\psi$ &  \multicolumn{3}{c} {$F-test$:value (status)$^\dag$ } & INV$^\dag$  & Ref. \\
               &       & (hr)     &   &   (mag)             &(mag)             & (mag)     & (1$e$-2\,mag)& (1$e$-2\,mag)&   (1$e$-2\,mag) & (t-s1) & (t-s2) & (s1-s2)     &    &     \\        
(1)       & (2) & (3) & (4) & (5)      & (6)      & (7)   & (8) & (9)  & (10) & (11) & (12) & (13)   & (14)    & (15)   \\ \hline
1998 Dec 26   & ST  & 7.8 & 49  & $-$0.531 & $-$0.407 & 0.124    &0.4  & 0.6  & 2.6  & 1.38(N)  & 1.05(N)  & 0.75(N)  &N & (a)   \\ 
1999 Dec 30   & ST  & 7.4 & 64  & $-$0.988 & $-$1.055 & $-$0.067 &0.4  & 0.5  & 2.1  & 0.49(N)  & 0.68(N)  & 0.62(N)  &N & (a)   \\
2000 Dec 24   & ST  & 6.0 & 42  & $-$1.676 & $-$1.737 & $-$0.061 &0.4  & 0.5  & 2.2  & 1.26(N)  & 0.70(N)  & 0.70(N)  &N & (a)   \\
2001 Dec 24   & ST  & 7.3 & 38  & $-$1.115 & $-$1.306 & $-$0.191 &0.3  & 0.4  & 1.8  & 1.46(N)  & 0.98(N)  & 0.52(N)  &N & (a)   \\
2003 Dec 20   & HCT & 6.0 & 38  & $-$0.774 & 0.045    & 0.819    & 0.2 & 0.3  & 1.5  & 1.76(PV) & 2.13(PV) & 0.80(N)  &PV& (b)   \\
2004 Dec 10   & ST  & 6.2 & 30  & 0.286    & $-$0.227 & $-$0.513 & 0.2 & 0.3  & 2.1  & 4.48(V)  & 3.54(V)  & 1.17(N)  &V & (b)   \\
2004 Dec 23   & ST  & 5.9 & 13  & 0.196    & $-$0.310 & $-$0.506 & 0.2 & 0.3  & 1.5  & 5.59(V)  & 3.35(PV) & 1.15(N)  &V & (b)   \\
2005 Jan 2     & ST  & 4.9 & 22  & 0.108    & $-$0.414 & $-$0.522 & 0.2 & 0.3  & 1.2  & 0.85(N)  & 1.65(N)  & 0.81(N)  &N & (b)   \\
2005 Jan 5     & ST  & 5.2 & 26  & 0.269    & 0.111    & $-$0.158 & 0.1 & 0.2  & 1.4  & 4.45(V)  & 3.09(V)  & 1.08(N)  &V & (b)   \\
2005 Jan 9     & ST  & 7.1 & 30  & 0.325    & 0.172    & $-$0.153 & 0.1 & 0.2  & 1.5  & 3.50(V)  & 3.66(V)  & 0.90(N)  &V & (b)   \\
2005 Nov 9    & ST  & 4.3 & 19  & 0.449    & 0.362    & $-$0.086 & 0.1 & 0.3  & 1.2  & 3.00(PV) & 2.32(PV) & 1.74(N)  &PV& (b)   \\
2006 Nov 16   & ST  & 5.0 & 21  & 0.580    & 1.550    & 0.970    & 0.2 & 0.4  & 1.4  & 1.72(N)  & 0.82(N)  & 2.49(N)  &N & (b)   \\
2006 Nov 29   & ST  & 6.5 & 28  & 0.947    & 0.430    & $-$0.517 & 0.2 & 0.3  & 1.9  & 1.31(N)  & 1.27(N)  & 1.00(N)  &N & (b)   \\
2006 Dec 17   & ST  & 6.5 & 28  & 1.013    & 0.505    & $-$0.507 & 0.2 & 0.3  & 2.2  & 1.87(N)  & 1.93(PV) & 1.45(N)  &N & (b)   \\
2007 Jan 11    & Ch  & 3.9 & 90  & 1.053    & 1.924    & 0.870    & 0.5 & 0.6  & 4.8  & 0.59(N)  & 0.58(N)  & 0.64(N)  &N & (c)   \\
2007 Dec 15   & ST  & 7.1 & 29  & 0.364    & 0.201    & $-$0.163 & 0.1 & 0.2  & 2.6  & 4.62(V)  & 4.99(V)  & 1.35(N)  &V & (b)   \\
2007 Dec 16   & ST  & 7.1 & 29  & 0.281    & $-$0.228 & $-$0.509 & 0.2 & 0.2  & 1.7  & 2.16(N)  & 1.31(N)  & 0.46(N)  &N & (b)   \\
2008 Nov 22   & ST  & 6.0 & 29  & 1.037    & 0.908    & $-$0.129 & 0.2 & 0.2  & 1.4  & 0.64(N)  & 0.74(N)  & 0.53(N)  &N & (b)   \\
2009 Jan 20    & ST  & 4.0 & 66  & 0.459    & 2.043    & 1.584    & 0.2 & 0.7  & 5.5  & 6.01(V)  & 7.47(V)  &5.26(V)   &N & (d)   \\
2009 Dec 8    & ST  & 6.9 & 31  & 0.455    & 0.326    & $-$0.129 & 0.4 & 0.5  & 3.2  & 1.65(N)  & 1.23(N)  & 0.91(N)  &N & (e)   \\
2011 Jan 5     & ST  & 6.8 & 32  & 1.284    & 0.953    & $-$0.331 & 0.3 & 0.4  & 3.8  & 1.13(N)  & 1.16(N)  & 0.51(N)  &N & (e)   \\
2011 Nov 29   & ST  & 6.1 & 29  & 1.113    & 0.613    & $-$0.500 & 0.2 & 0.3  & 1.9  & 1.18(N)  & 0.65(N)  & 0.51(N)  &N & (e)   \\
2012 Dec 20   & ST  & 6.9 & 115 &          &          &          &     &      &      &            &            &      &N & (f)   \\
2013 Jan 7     & ST  & 5.6 & 22  & 0.535    & 0.383    & $-$0.152 & 0.3 & 0.4  & 4.9  & 6.43(V)  & 7.00(V)  & 0.91(N)  &V & (e)   \\
2015 Feb 15   & AOJU& 4.1 & 30  & 0.630    & 1.143    & 0.513    & 0.8 & 1.1  & 9.1  & 1.99(PV) & 1.86(PV) & 0.81(N)  &PV& (e)   \\
\hline
\end{tabular}
\begin{minipage}{\textwidth}
Columns: (1) date of observation; (2) telescope used; (3) duration of monitoring; (4) number of data points in the DLC; (5) mean magnitude difference of the t-s1 DLC; (6) mean magnitude difference of the t-s2 DLC; (7) mean magnitude difference of the s1-s2 DLC; (8) quadratic mean of the $\sc IRAF$ errors for the s1-s2 DLC; (9) standard deviation of the s1-s2 DLC; (10) INV amplitude ($\psi$); (11) $F-$ value obtained for the t-s1 DLC (variability status of the DLC); (12) $F-$ value obtained for the t-s2 DLC (variability status of the DLC); (13) $F-$ value obtained for the s1-s2 DLC (variability status of the DLC); (14) Variability status of BL Lac; (15) Reference for the INV data : (a) \citet{2004MNRAS.348..176Sagar}; (b) \citet{2009MNRAS.399.1622AG}; (c) \citet{2008AJ....135.1384Gupta}; (d) \citet{2011MNRAS.413.2157Rani}; (e) present work; (f) \citet{2012MNRAS.424.2625B};  
$^\dag$ V = Variable; N = Non-variable; PV = Probable Variable;
\end{minipage}
\end{table*}

\begin{figure*}[h!]
\centering
\includegraphics[width=0.41\textwidth]{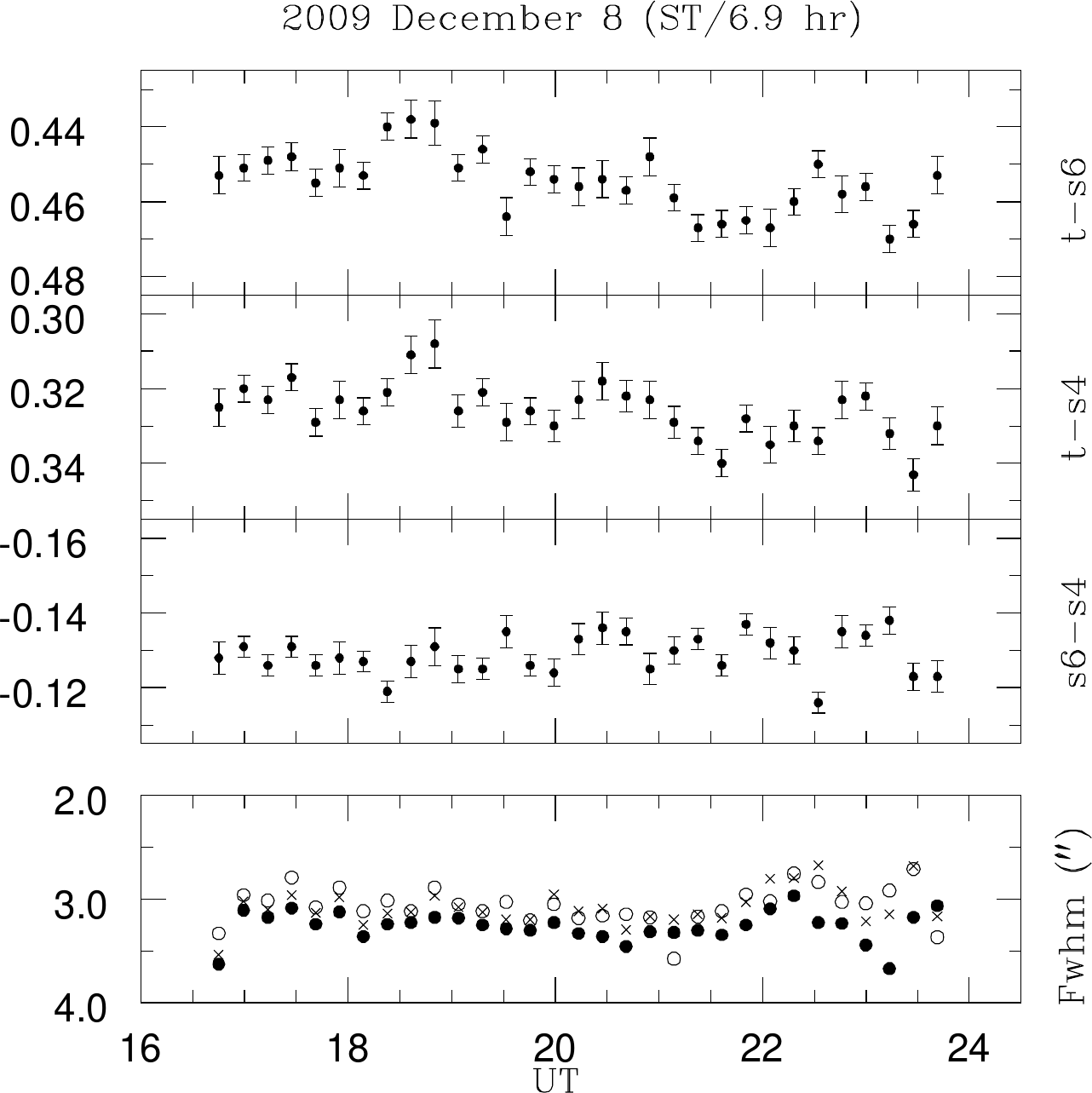}
\includegraphics[width=0.41\textwidth]{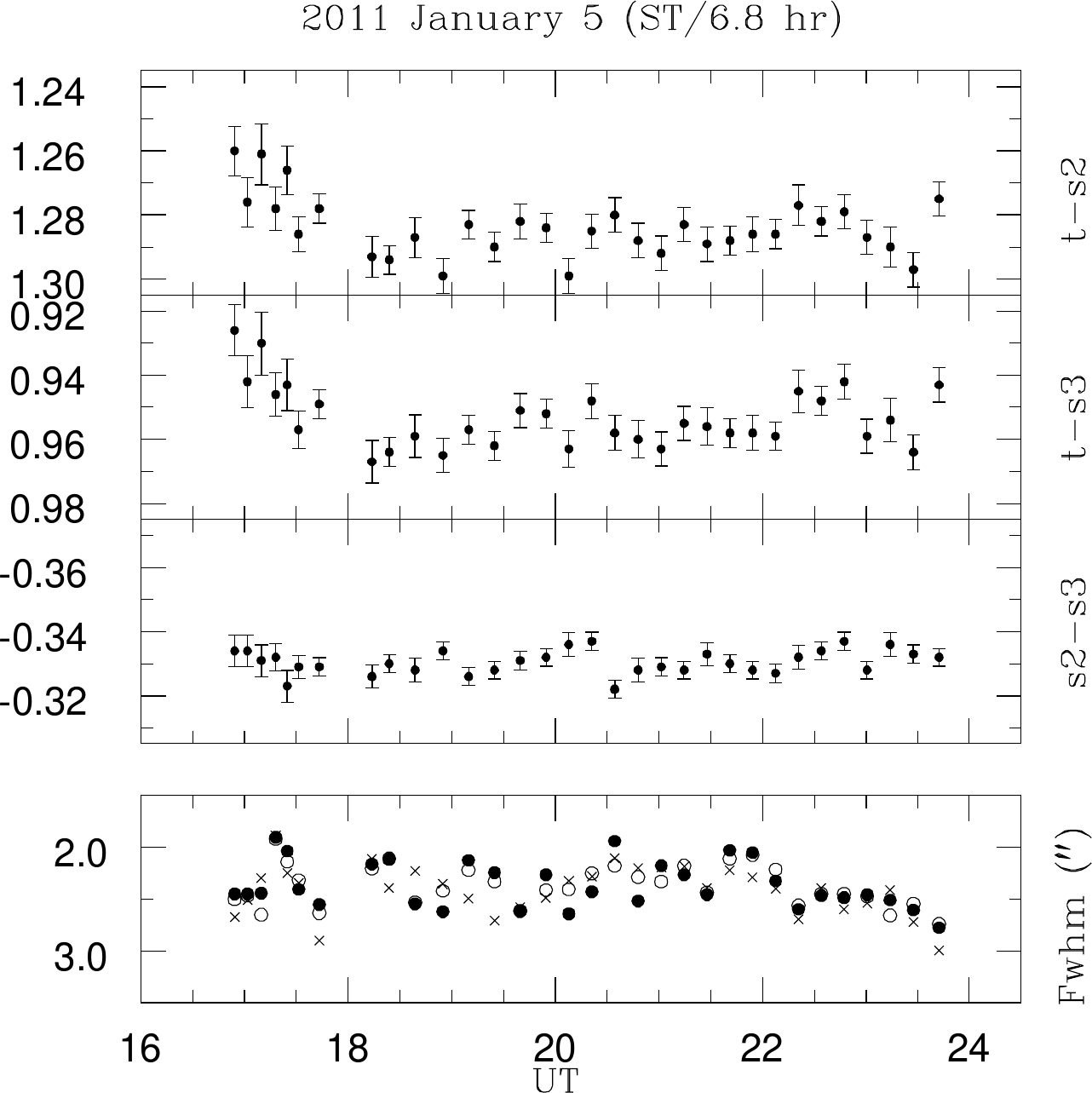}
\includegraphics[width=0.41\textwidth]{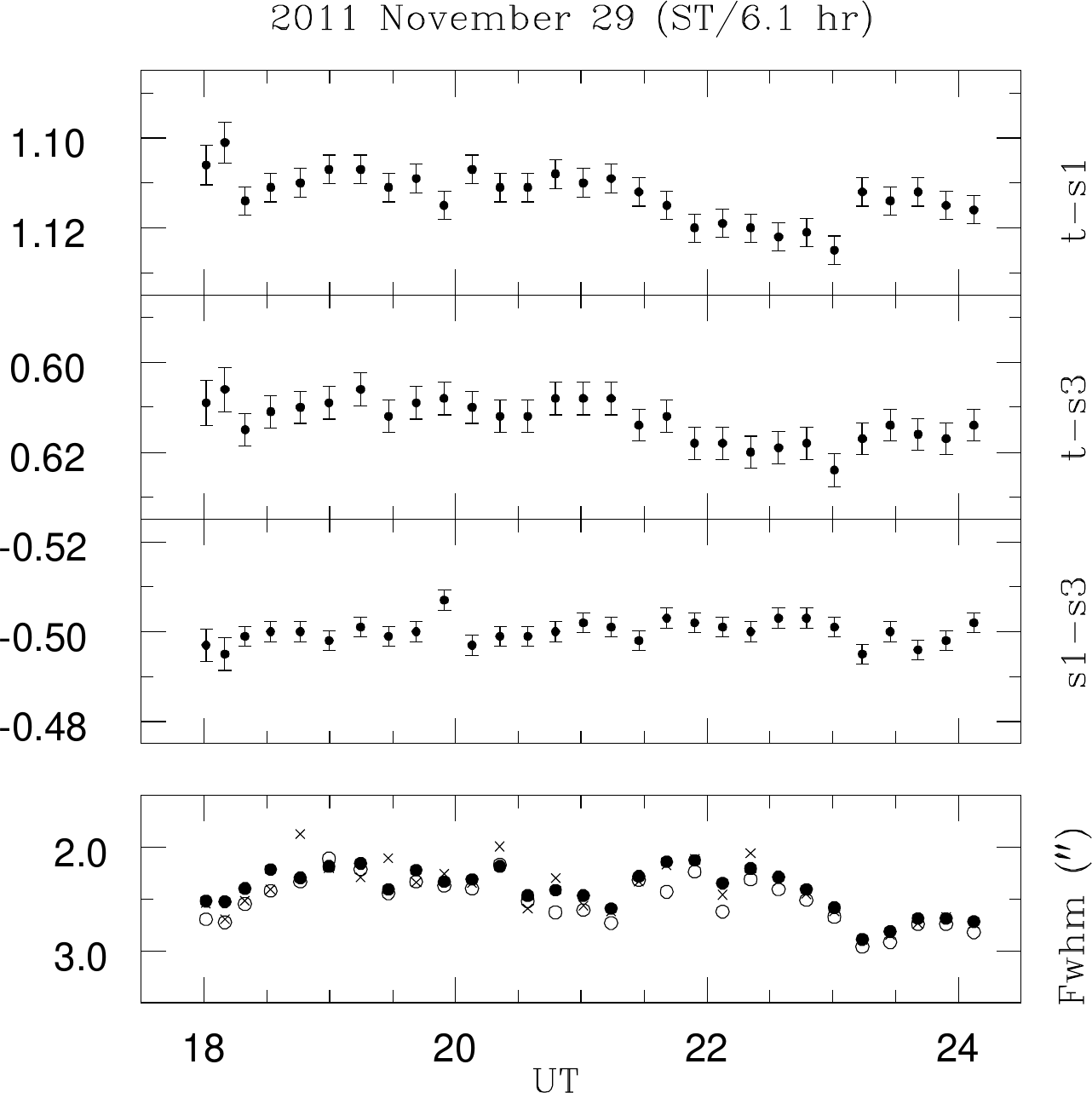}
\includegraphics[width=0.41\textwidth]{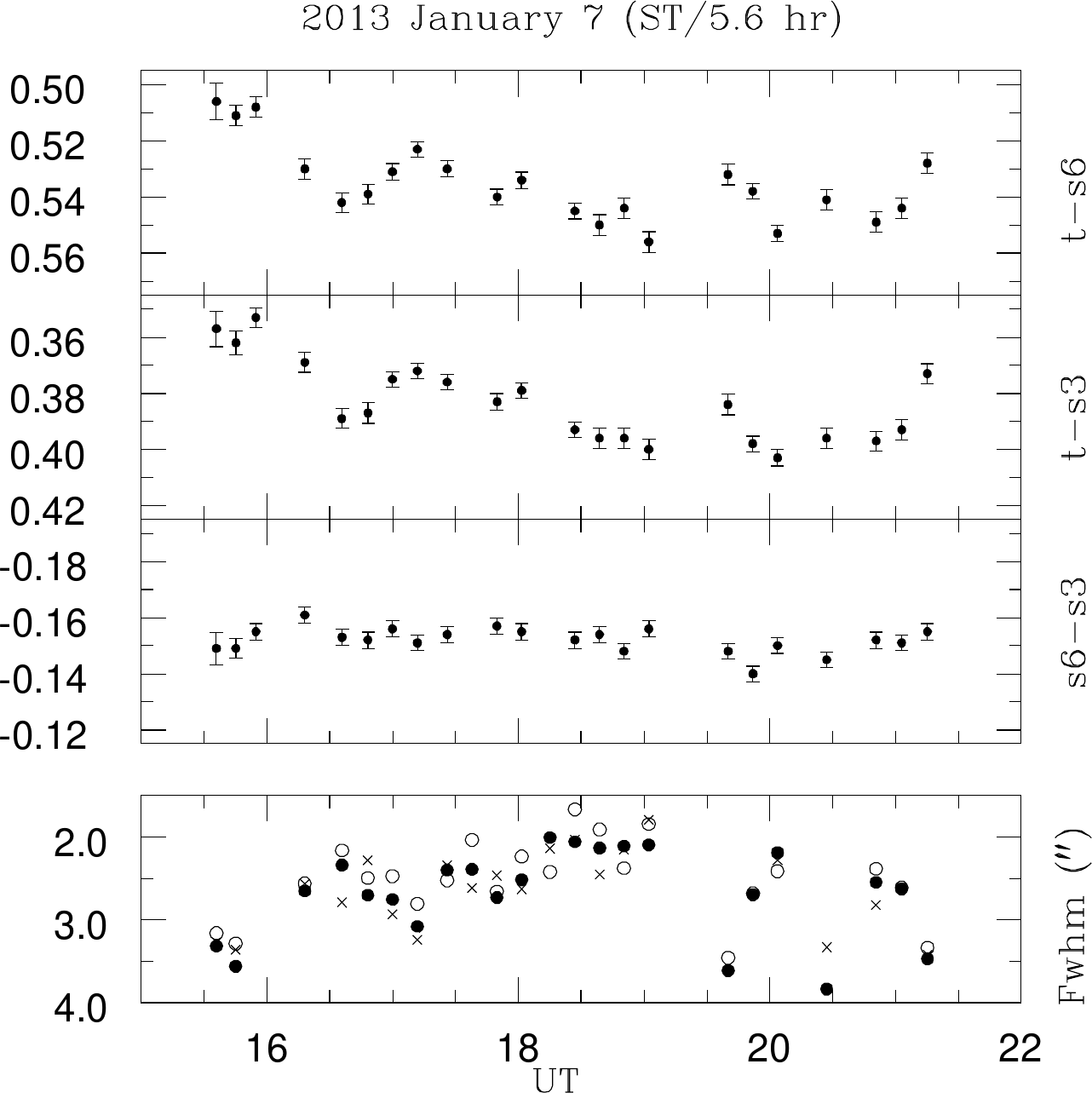}
\includegraphics[width=0.41\textwidth]{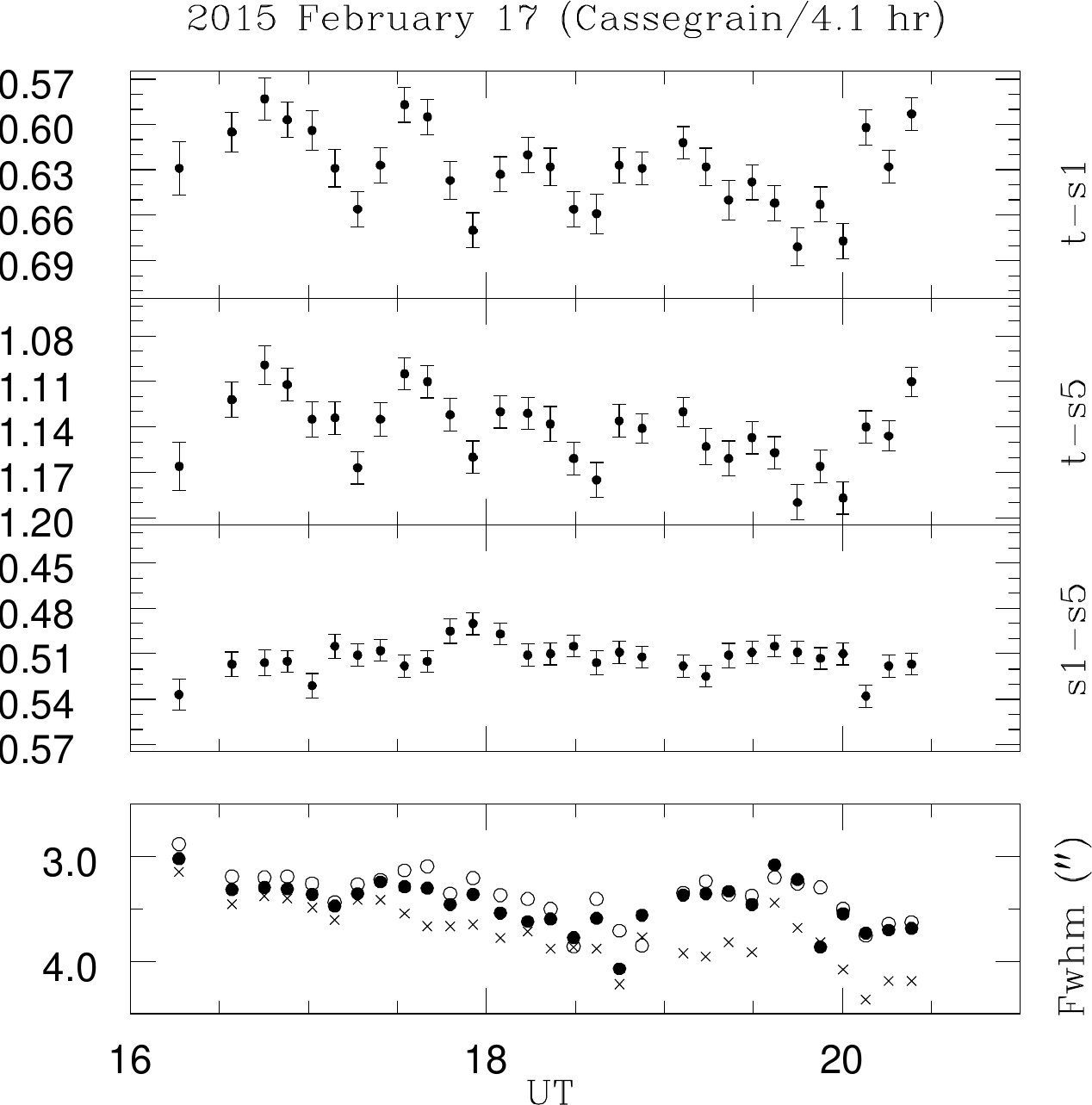}
\caption{Our 5 newly acquired R-band intra-night optical DLCs of the BL Lac object PKS\,0735$+$178. The date, the telescope used, and the duration of monitoring are given at the top of each nights' plots. The upper two panels show the DLCs of the target relative to two steady comparison stars, while the third panel shows the star-star DLC. The bottom panels give the plots of seeing variation for the night, based on the three stars (shown by crosses, open circles, and filled circles, respectively), monitored along with the target blazar on the same CCD frame.}
\label{fig:INV}
\end{figure*}

As revealed by  very long baseline interferometry (VLBI) imaging, the radio jet of the blazar underwent several dramatic structural changes between 1981 and 2001 on the scale of 10\,pc, altering between a `staircase jet' with a highly bent trajectory (circa 1981 and 1995), and a straight jet with a linear trajectory (1985 and 2001); the last morphological transition coincided with the optical flaring but was accompanied by only a mild increase in the radio intensity \citep{2010A&A...515A.105B}. By inspecting milli-arcsec scale resolution images from the MOJAVE database\footnote{\scriptsize{\texttt{http://www.physics.purdue.edu/astro/MOJAVE/sourcepages/0735+178.shtml}}} and the Boston University database\footnote{\scriptsize{\texttt{https://www.bu.edu/blazars/VLBA\_GLAST/0735.html}}}, we confirmed that the jet trajectory has remained linear from 2001 until 2015. Faraday rotation gradients and circular polarization have been detected in the radio core of PKS\,0735$+$178 \citep{2008MNRAS.384.1003G}. The jet magnetic field structure revealed by the radio polarization maps is rather complex (though again, not unusual for a low-frequency-peaked BL Lac), consisting of magnetic field lines mostly perpendicular to the jet axis near the core region, but parallel to the jet further along the outflow; this change suggests either a helical configuration with changing pitch angle or strong interactions with the ambient medium leading to velocity shears and jet bends \citep{2006A&A...453..477A}. 

In Section 2 we describe our data and its reduction. Our analysis and results are given in Section 3 and Section 4 provides a discussion of the results and our main conclusions 

\section{Data acquisition}
\label{sec:obs}

\subsection{Optical: intra-night}

The vast majority of the intra-night observations were carried out using the 104-cm Sampurnanand telescope (ST) located at the Aryabhatta Research Institute of observational sciencES (ARIES), Naini Tal, India. The ST has Ritchey-Chr\'etien (RC) optics with a f$/$13 beam \citep{1999CSci...77..643G}. The detector was a cryogenically cooled $2048 \times 2048$ chip mounted at the Cassegrain focus. This chip has a readout noise of 5.3\,e$^{-}$/pixel and a gain of 10\,e$^{-}$$/$Analog to Digital Unit (ADU) in slow readout mode. Each pixel has a dimension of 24\,$\mu$m$^{2}$ which corresponds to 0.37\,arcsec$^{2}$ on the sky, covering a total field of $13^{\prime} \times 13^{\prime}$. Our observations were carried out in $2 \times 2$ binned mode to improve the signal-to-noise ratio.

We similarly used the 201-cm Himalayan Chandra Telescope (HCT) at the Indian Astronomical Observatory (IAO), located in Hanle, India. This telescope is also of the RC design but has a f$/$9 beam at the Cassegrain focus\footnote{\texttt{http://www.iiap.res.in/$\sim$iao}}. The detector was a cryogenically cooled $2048 \times 4096$ chip, of which the central $2048 \times 2048$ pixels were used. The pixel size is 15\,$\mu$m$^{2}$, so that the image scale of 0.29\,arcsec$/$pixel covers an area of $10^{\prime} \times 10{^\prime}$ on the sky. The readout noise of CCD is 4.87\,e$^{-}$/pixel and the gain is 1.22\,e$^{-}$$/$ADU. The CCD was used in the unbinned mode.

Lastly, we also employed the 50-cm Cassegrain telescope  in the Astronomical Observatory of the Jagiellonian University (AOJU), located in Krak\'ow, Poland. This telescope is also of the RC design with f$/$6.7 beam at the Cassegrain focus. The detector was a thermoelectric cooled $1024 \times 1024$ chip, corresponding to an image scale of 0.70\,arcsec$/$pixel, covering a total of $12^{\prime} \times 12{^\prime}$ on the sky. The CCD was used in $2 \times 2$ binned mode. 
 
The seeing mostly ranged between $\sim 1^{\prime\prime}.5$ to $\sim 3^{\prime\prime}$, as determined using three sufficiently bright stars on each CCD frame. All the observations were made using the R filter, as the CCD responses are maximized in this band. The field positioning was adjusted in order to have within each CCD frame at least two, and usually three, comparison stars. For all the telescopes, bias frames were taken intermittently, and twilight sky flats were also obtained.

The pre-processing of the images (bias subtraction, flat-fielding and cosmic-ray removal) was done by applying the standard procedures in the Image Reduction and Analysis Facility ({\sc IRAF})\footnote{\texttt{http://iraf.noao.edu/}} software package. The instrumental magnitudes of the target AGN and the stars (all point-like) in the image frames were determined by the aperture photometry using {\sc APPHOT}. The magnitude of the target AGN was measured relative to a few apparently steady comparison stars present on the same CCD frame; relative star-star magnitudes were also recorded. In this way, the Differential Light Curves (DLCs) for the AGN and the comparison stars were derived. Out of the resulting three star-star DLCs, we selected the steadiest star-star DLC (based on the lowest variance) for testing the INV of the blazar monitored on a given night. These chosen stars are hereafter named `s1' and `s2', and the corresponding target-star and star-star DLCs are denoted as `t-s1', `t-s2', `s1-s2', respectively. Basic information about the  comparison stars (apparent magnitudes, optical colors, positions on the sky) is given in \citet{2009MNRAS.399.1622AG}.

The comparison stars we used are typically within about one magnitude of the target AGN; note that avoiding large differences in brightness is of crucial importance for minimizing the possibility of a spurious INV detection \citep[e.g.,][]{2007MNRAS.374..357Cellone}. In this context, spurious variability on account of different second-order extinction coefficients for the AGN and their comparison stars can also be problematic if the target AGN and the comparison stars have very different optical colors. However, as shown by \citet{1992AJ....104...15Carini} and \citet{2004JApA...25....1Stalin}, for color differences of up to 1-2 mag, the differential extinction of photons traveling through varying airmasses do not influence significantly the derived INV parameters of BL Lac objects, given their typical flux measurement uncertainties ($\simeq 0.1-0.2\%$, including the observations analyzed here). For each night, an optimum aperture radius for the photometry was chosen by identifying the minimum dispersion in the star-star DLC, starting from the median seeing (i.e., full width at half maxima) value on that night to four times that value. For a given night, we  selected the aperture size which yielded the minimum dispersion in the steadiest star-star DLC. This also set the threshold for the INV detection on that night \citep[see][]{2013JApA...34..273Goyal}. Typically, the selected aperture radius was $\sim 4^{\prime\prime}$ and the effective seeing was $\sim 2^{\prime\prime}$.

Our entire intra-night data are summarized in Table~\ref{tab:result}, and the newly acquired DLCs are shown in Figure~\ref{fig:INV}. 

\subsection{Optical: long/short-term}

The nightly averaged R-band photo-polarimetric data for the period 2005 October 22 to 2015 September 27 were obtained using (i) the 70-cm AZT-8 telescope of the Crimean  Astrophysical  Observatory (Nauchnij, Ukraine), (ii) the 40-cm LX-200 telescope of the St. Petersburg State University (St. Petersburg, Russia), (iii) the 1.8-m Perkins  telescope  of the Lowell Observatory (Flagstaff AZ, USA), (iv) the 1.54-m Kuiper and 2.3-m Bok telescopes of the Steward Observatory (Mt.\ Bigelow AZ and Kitt Peak AZ, USA), and (v) the 2.2-m telescope of the Calar Alto Observatory (Calar Alto, Spain) within the MAPCAT program\footnote{\texttt{http://www.iaa.es/$\sim$iagudo/research/MAPCAT/MAPCAT.html}} (see \citealt{2008A&A...492..389L} for the description of the program and analysis). The photometric data were supplemented with the nightly averaged R-band optical data from 1993 until 2005, obtained from \citet{2007A&A...467..465C}. These nightly averaged data have quoted photometric uncertainty of the order of $\sim 5-10\%$, arising mainly from large calibration errors in the estimated magnitudes of the stars in the field  \citep{2007A&A...467..465C}. It is a standard procedure to use only one or two stars to scale the magnitude of a blazar to the standard system. In case of intra-night data, however, as a very high photometric accuracy is required to detect variability down to $1\%$ amplitudes, we standardized our comparison stars \citep[Table 2 of][]{2009MNRAS.399.1622AG} using all the standard stars in the field of the target, as listed in \citet{2007A&A...467..465C}; this resulted in the accuracy of $\leq 0.2-0.5\%$. Finally, for a given R-band magnitude $M_{\rm R}$ the R-band flux (in Jy) was derived as $3064 \times 10^{-0.4\times M_{\rm R}}$, where 3064\,Jy is the zero point magnitude flux of the photometric system \citep{1999hia..book.....G}; the errors in R-band fluxes were derived using standard error propagation \citep{2003drea.book.....Bevington}. 
 
\subsection{High energy $\gamma$-rays: long-term} 
\label{sec:fermi}

We have analyzed the {\it Fermi}-LAT data for the field containing PKS\,0735+178 from 2008 August through 2015 September, and produced a source light curve between 0.1 and 200\,GeV with an integration time of 15 days. We have performed the unbinned likelihood analysis using Fermi ScienceTools-v10r0p5 with {\sc p8r2\_source\_v6} source event selection and instrument response function, for the $20 ^\circ$ region centered at the blazar, following the Fermi tutorial\footnote{\texttt{http://fermi.gsfc.nasa.gov/ssc/data/analysis/scitools/}}. The procedure starts with the selection of good data and time intervals (using the tasks {\sc `gtselect'} and {\sc `gtmktime'} with selection cuts {\sc evclass=128 evtype=3}), followed by the creation of an exposure map in the region of interest (ROI) with $30^\circ$ radius for each time bin (tasks {\sc `gtltcube'}, {\sc `gtexpmap'} while counting photons within zenith angle $< 90^\circ$).  We then computed the diffuse source response (task {\sc `gtdifrsp'}), and finally modeled the data through a maximum-likelihood method (task {\sc `gtlike'}). In this last step, we used a model that includes PKS\,0735+178 and 170 other sources inside the ROI \citep[according to the third Fermi Large Area Telescope source catalog, 3FGL;][]{2015ApJ...810...14A}. The model also takes into account the diffuse emission from our Galaxy and the extragalactic $\gamma$-ray background\footnote{{\sc gll\_iem\_v06.fits} and {\sc iso\_p8r2\_source\_v6\_v06.txt}}\citep{2016ApJS..223...26A}. In the modelling, we followed the usual method and fixed the spectral indices and fluxes of all the point sources within the ROI,  other than the target, at their 3FGL values. The $\gamma$-ray spectrum of PKS\,0735+178 was modeled with a simple power law. We considered a measurement to be a successful detection for a test statistic TS\,$\geq$\,10, which corresponds to a signal-to-noise ratio $\geq 3\sigma$ \citep{2009ApJS..183...46A}.

\begin{figure*}[h!]
\centering
\includegraphics[width=\textwidth]{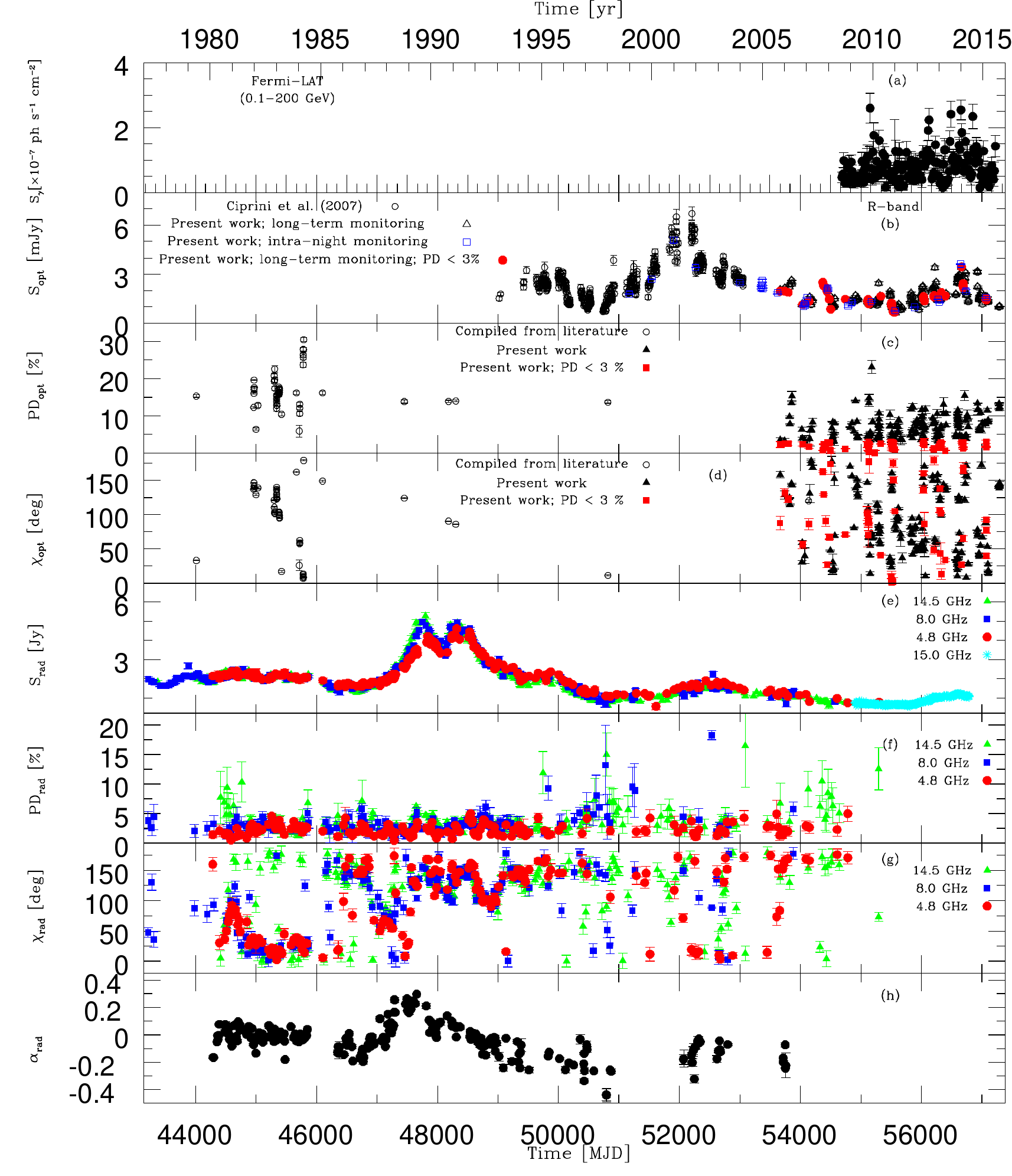}
\caption{The multiwavelength, photo-polarimetric, and long-term variability light curves of PKS\,0735+178. Panel (a) shows the {\it Fermi-}LAT light curve at energy range 0.1-200 GeV. Panels (b--d) show the R-band total flux, polarization degree (PD) and electric vector position angle ($\chi$). Panels (e--g) show the GHz band total radio flux, PD and $\chi$ while, panel {\bf(h)} shows the run of radio spectral index obtained at GHz frequencies.}
\label{fig:LC}
\end{figure*}

\clearpage

\subsection{Radio: long-term}

The radio data were obtained from the University of Michigan Radio Astronomy Observatory (UMRAO) 26m dish at 4.8, 8.0, and 14.5\,GHz, and the 40-m Telescope at the Owens Valley Radio Observatory (OVRO) at 15\,GHz. The UMRAO fluxes at 4.8, 8.0, and 14.5\,GHz were typically sampled twice per month \citep{1985ApJ...298..296A}, from 1980 February 16 -- 2010 April 16, from 1977 March 11 -- 2008 April 5, and from 1977 August 19 -- 2011 January 24, respectively, while the OVRO light curve at 15\,GHz was sampled twice a week \citep{2011ApJS..194...29R}, during the period from 2009 march 17 to 2014 May 8.  The discussion on the corresponding observing strategy and calibration procedures can be found in \citet{1985ApJ...298..296A} for the UMRAO data, and in \citet{2011ApJS..194...29R} for the OVRO data. 
 
\section{Data analysis and Results}

\subsection{Optical microvariability}
\label{sec:micro}

In our analysis, we used the $F-$test for assigning the INV detection significance. The $F-$statistic compares the observed variance $V_{\rm obs}$ to the expected variance $V_{\rm exp}$. The null hypothesis of no variability is rejected when the ratio
\begin{equation}
F_\nu^\alpha =\frac{V_{\rm obs}}{V_{\rm exp}} = \frac{V_{t-s}}{\langle \eta^2 \, \sigma_{t-s}^2 \rangle} ,
\label{eq:ftest}
\end{equation}
exceeds a critical value for a chosen significance level $\alpha$, for a given number of degrees of freedom (DOF) $\nu$; here $V_{t-s}$ is the variance of the `target-star' DLC, $\langle \sigma_{t-s}^2 \rangle$ is the mean of the squares of the (formal) rms errors of the individual data points in the `target-star' DLC. Since this method requires flux or magnitude estimates along with their error estimates, it is important to determine the photometric errors accurately. As emphasized in several independent studies, the photometric errors returned by $APPHOT$ are significatnly underestimated (\citealt{2004JApA...25....1Stalin} and references therein). \citet{2013JApA...34..273Goyal}  reported the latest attempt to determine this under-estimation factor, $\eta$, using an unprecedented data set consisting of 262 steady star-star DLCs. They find $\eta= 1.54\pm0.05$ which confirms the previous estimates by the same group which were based on much smaller data sets. \citet{2013JApA...34..273Goyal} also showed that the determination of $\eta$ is quite insensitive to the magnitude difference   between the pair of objects used for deriving a DLC, as long as it did not exceed 1.5 mag. Thus, $\eta=$1.54 has been used in the present analysis to scale up the {\sc IRAF} photometric magnitude errors (see also \citealt{1995MNRAS.274..701G, 2010ApJ...723..737Villforth}). Note that the  standard expression for $F$ is given by $F_{\nu_1,\nu_2}^{\alpha} = \sigma_1^2/\sigma_2^2 $, where $\sigma_1$ and $\sigma_2$ are the two variances with the corresponding DOF, $\nu_1$ and $\nu_2$. In our analysis, we have simplified this definition since $\nu_1$ = $\nu_2 = \nu$ is also the number of DOF for the `star-star' DLC.

The significance level set for a given test determines the {\it expected} number of {\it false positives}, which is an indicator of the robustness of the test. We have chosen two significance levels, $\alpha = $ 0.01 and 0.05, corresponding to $p-$values of $\ga$ 0.99 and $\ga$ 0.95, respectively. Recall that the smaller the value of $\alpha$ is, the less likely it is for the variability to occur by chance. Thus, in order to claim a genuine INV detection, i.e., to assign a `variable' designation (V), we stipulate that the computed statistic value is above the critical value corresponding to $p > 0.99$ (i.e., $\alpha=$ 0.01) for a given degree of freedom ($\nu = N_p - 1$, where $N_p$ stands for the number of data points in a given DLC). We assign a `probable variable' designation (PV) when the computed test statistic value is found to be between the critical values at $\alpha = $ 0.01 and 0.05; otherwise, a `non-variable' (N) designation is assigned to a DLC. All the three DLCs, i.e., $t-s1, t-s2$ and $s1-s2$, are subjected to the $F-$test analysis. In a few cases, the INV status was different for the two blazar-star DLCs, and this indicated a small amplitude variation of one or the other comparison stars. Since such small amplitude variations in star-star DLCs are difficult to ascertain, we only ascribed a ``V'' status if both blazar-star DLCs gave a ``V'' status; otherwise, we quote a``PV'' or ``N'' status if the star-star DLC itself turned to be variable. The analysis results are summarized in Table~\ref{tab:result}.

Following \citet{1999A&AS..135..477Romero} the peak-to-peak INV amplitude was calculated as 
\begin{equation} 
\psi= \sqrt{({D_{\rm max}}-{D_{\rm min}})^2-2\sigma^2} \, ,
\label{eq:psi}
\end{equation}
with $D_{\rm min/max}$ denoting the minimum/maximum in the differential light curve of the source, and $\sigma^2= \eta^2 \, \langle\sigma^2_{i}\rangle$ where $\eta =1.54$ and $\sigma_i$ is the nominal error associated with each data point. 

The INV DC was computed according to
\begin{equation} 
DC = 100\% \,\,\, \frac{\sum_{j=1}^n N_j \, (1/\Delta t_j)}{\sum_{j=1}^n (1/\Delta t_j)} \, ,
\label{eq:DC}
\end{equation}
where $\Delta t_j = \Delta t_{j,\, {\rm obs}} \, (1+z)^{-1}$ is the duration of the monitoring session of a source on the $j^{th}$ night, corrected for the cosmological redshift $z$, and $N_j$ is set equal to 1 if INV was detected, and otherwise to 0 \citep{1999A&AS..135..477Romero,2004JApA...25....1Stalin}. This estimate is essentially the ratio of the number of nights a source is found to be variable to the total number of nights it was monitored. Note that, since for a given source the monitoring times on different nights are not necessarily equal, the evaluation of the DC has been appropriately weighted by the actual monitoring duration $\Delta t_j$. The computed INV DC for the entire data set, consisting of 25 nights spanning over 18 years (1998--2015) is 22\% (37\% if three PV cases are included). The INV DC for the stronger nightly variations,  where $\psi > 3\%$, is 4\% (10\% if one PV case is included). 

In order to validate our analysis procedure, we have performed a sanity check by computing the number of `Type 1 errors', or the false positives, for our data set. A false positive arises due the rejection of a true null hypothesis by a test, when applied to a non-varying DLC (i.e., the inability to discern a non-variable object as non-variable). Assuming {\it a priori} that the star-star DLCs are steady, the outcome of the statistical test applied to these should be consistent with the {\it expected} number of false positives for the assumed value of $\alpha$. The number of  false positives depends only on the number of the star-star DLCs examined and the value of $\alpha$ chosen for the test. Thus, if the number of false positives is found to be significantly different from the expected number, either the test is not robust or the the measurement errors are not properly accounted for \citep[see, e.g.,][for the robustness of statistical tests in microvariability studies]{2010AJ....139.1269Diego,2013MNRAS.435.1300AG}. 

We note that for our data set consisting of 25 steady star-star DLCs, the means of the {\it expected} numbers of false positives are $\simeq 0.3$ and $\simeq 1.3$ for $\alpha= 0.01$ and 0.05, respectively. Since the distribution of false positives is expectedly binomial, for $\alpha = 0.01$ the number of false positives should in fact be scattered between 0 and 2, and for most of the cases around $\simeq 0.3\pm 0.5$. Similarly, with $\alpha = 0.05$, the number of false positives should lie between 0 and 5, and should largely cluster at $\simeq 1 \pm 1$. Meanwhile, the {\it observed} numbers of false positives reported by the application of the $F-$test (see column~13 of Table~\ref{tab:result}) is 1 for $\alpha = 0.01$ and 1 for $\alpha = 0.05$. The good agreement between the {\it expected} and the {\it observed} numbers of false positives provides validation for our analysis procedure.    

\subsection{Multi-wavelength long-term variability}

\begin{figure}[t!]
\centering
\includegraphics[width=\columnwidth]{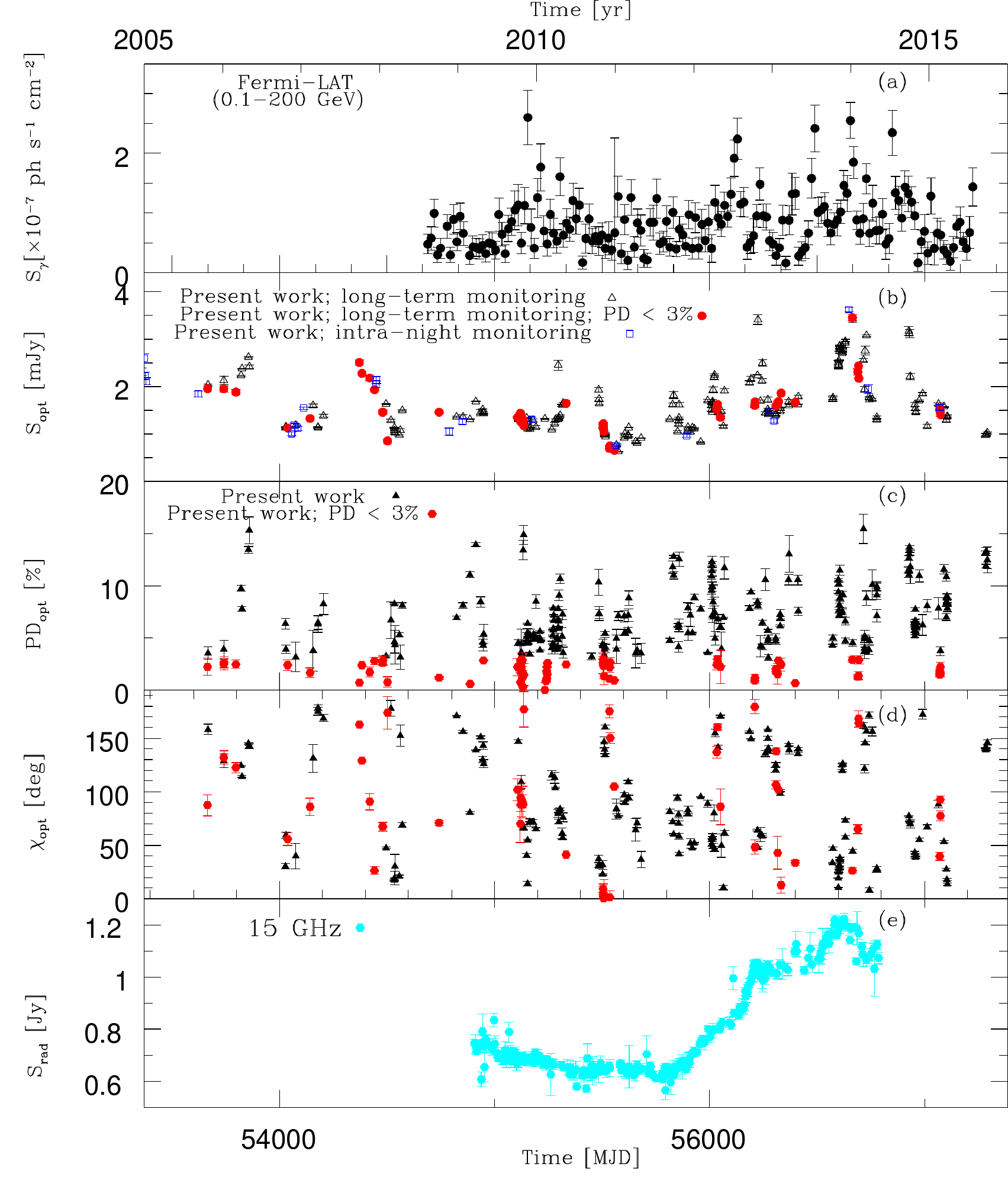}
\caption{An expanded view of the $\gamma-$ray, R-band photo-polarimetric, and radio long-term variability light curves of PKS\,0735$+$178 for the period 2005--2015.}
\label{fig:ZLC}
\end{figure}

Figure~\ref{fig:LC} presents the long-term, multi-wavelength, and photo-polarimetric light curve of PKS\,0735$+$178. Note that a brief discussion of the long-term optical and radio variability of the source (until 2008) was given in \citet{2009MNRAS.399.1622AG}. The figure includes the 0.1--200\,GeV integrated $\gamma$-ray flux $S_{\gamma}$ (panel 2a), the optical R-band photometric flux $S_{\rm opt}$, polarization degree ${\rm PD_{opt}}$, and polarization angle $\chi_{\rm opt}$ (panels 2b, 2c, and 2d, respectively), as well as the radio $S_{\rm rad}$, ${\rm PD_{rad}}$, $\chi_{\rm rad}$, and the spectral index $\alpha_{\rm rad}$ (panels 2e--2h). The radio spectral index, defined here as $S_{\nu}\propto{\nu^{\alpha}}$, was calculated by a linear regression analysis of the flux values at the three GHz frequencies in log-log space (with measurements at various radio frequencies performed within 14 days  considered as simultaneous). The optical polarization measurements (prior to 2005) were obtained from \citet{1987ApJS...64..459S}, \citet{1998AJ....116.2119V}, \citet{2009MNRAS.397.1893V}, and \citet{2011ApJS..194...19W}. Figure~\ref{fig:ZLC} shows the expanded long-term light curve consisting of our newly acquired data for the period 2005--2015.

\subsubsection{Power spectral analysis}
\label{sec:PSD}

The PSDs of the observed light curves of PKS\,0735+178 were generated using the standard methods outlined in \citet{2002MNRAS.332..231U} and \citet{2014MNRAS.445..437M}, which can be summarized as follows.  The PSD of an evenly sampled light curve $f(t_i)$ with mean $\mu$ and a total duration of $T$, consisting of $N$ data points, is defined as the rms-normalized periodogram,
\begin{equation}
P(\nu_k) = \frac{2 \, T}{\mu^2 \, N^2} \, | F(\nu_k) |^2 \, ,
\label{eq:psdeq}
\end{equation}
where $| F(\nu_k) |^2$ is the squared modulus of the discrete Fourier transform (DFT) calculated after subtracting the mean from flux measurements. In order to achieve a regular sampling of the light curve at evenly spaced time intervals, when our data were necessarily obtained at irregular intervals, we linearly interpolate between the two consecutive observed data points on the timescales typically 15-20 times smaller than the original observed sampling interval.  Reasons for subtracting the mean and performing data interpolation in the case of colored noise-type sources are discussed  in  Appendix\,\ref{sec:App}.

By definition, the integration of the periodogram over positive frequencies yields the total excess variance. Meanwhile, the noise floor levels corresponding to the variability power due solely to statistical fluctuations are estimated following \citet{2015ApJ...798...27I} as
\begin{equation}
P_{stat} = \frac{2 \, T}{\mu^2 \, N} \, \sigma_{\rm stat}^2 \, .
\label{eq:poi_psd}
\end{equation}
In the above, $\sigma_{stat}^2= \sum_{j=1}^{j=N} \Delta f(t_j)^2 / N$ is the mean variance of the measurement uncertainties on the flux values $\Delta f\!(t_j)$ in the observed light curve at times $t_j$, with  $N$ denoting the  number of data points in the original light curve. In the case of Gaussian errors, the noise floor level is scaled using typical sampling intervals of 1 day for the long term optical light curve, 3 days for the OVRO light curve, and 15 days for the UMRAO light curves \citep[see also][Appendix A]{2003MNRAS.345.1271V}. Note that in the case of the lower frequency UMRAO data, in particular the 4.8\,GHz ones, the calculated noise floor level is relatively high when compared with the corresponding variability power (see Fig.\,\ref{fig:radio}, bottom panel), and is not obviously reflected as a high-frequency plateau in the raw periodogram; this is solely due to the fact that in this case the errors on flux measurements were effectively over-estimated during the first quarter of the survey program.

\begin{figure}[t!]
\centering
\includegraphics[width=\columnwidth]{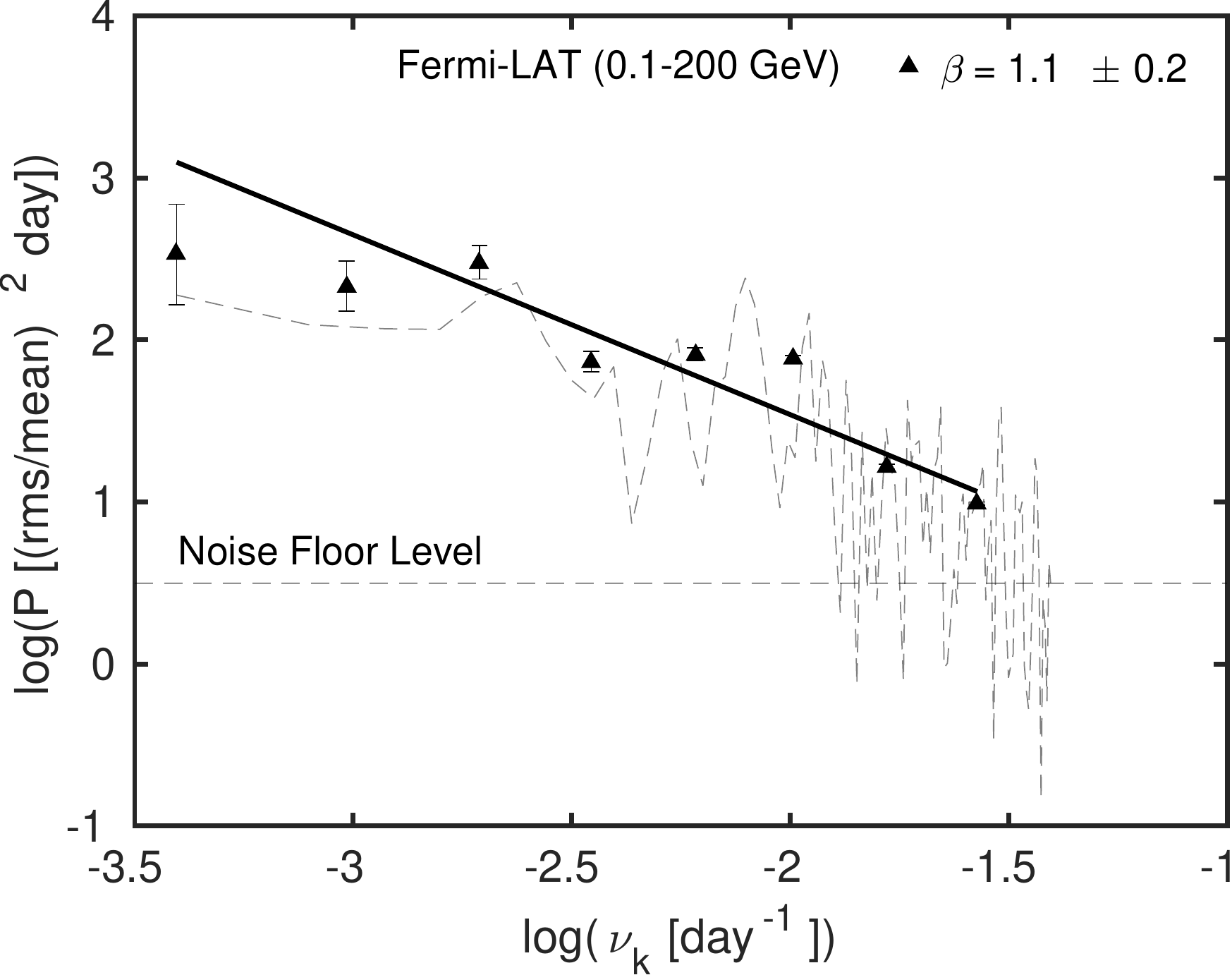}
\caption{The high-energy $\gamma$-ray ({\it Fermi}-LAT) PSD of PKS\,0735+178, corresponding to the data analyzed in this paper.
The dashed gray line and filled upward pointing traingles denote the raw and binned periodogram estimates, respectively, while the dashed horizontal line indicates the noise floor level due to the measurement error achieved. The solid black line is the least-squares fit to the binned periodogram.}
\label{fig:gamma}
\end{figure}

PSDs generated using this method are distorted due to the effects of the discrete sampling and the finite duration of the light curve, known as {\it aliasing} and  {\it red noise leakage}, respectively. The net effect of aliasing is to fold back the power from high frequencies to lower frequencies; at the same time, the red noise leak causes extra power at higher frequencies through the side-lobes of a sampling window function \citep{1992nrca.book.....Press}. The aliasing is reduced for steeper power-law power spectra \citep{1993MNRAS.261..612P, 2002MNRAS.332..231U}, while the red-noise leak can be minimized using a proper sampling window function \citep[see][for a detailed discussion]{2014MNRAS.445..437M}. In our analysis of the PSD, we used the `Hanning' window function as it has the lowest side lobes and the fastest fall-off rate as a function of frequency when compared to other frequently employed window functions (see Appendix\,\ref{sec:App} for a detailed discussion).

\begin{figure}[t!]
\centering
\includegraphics[width=0.83\columnwidth]{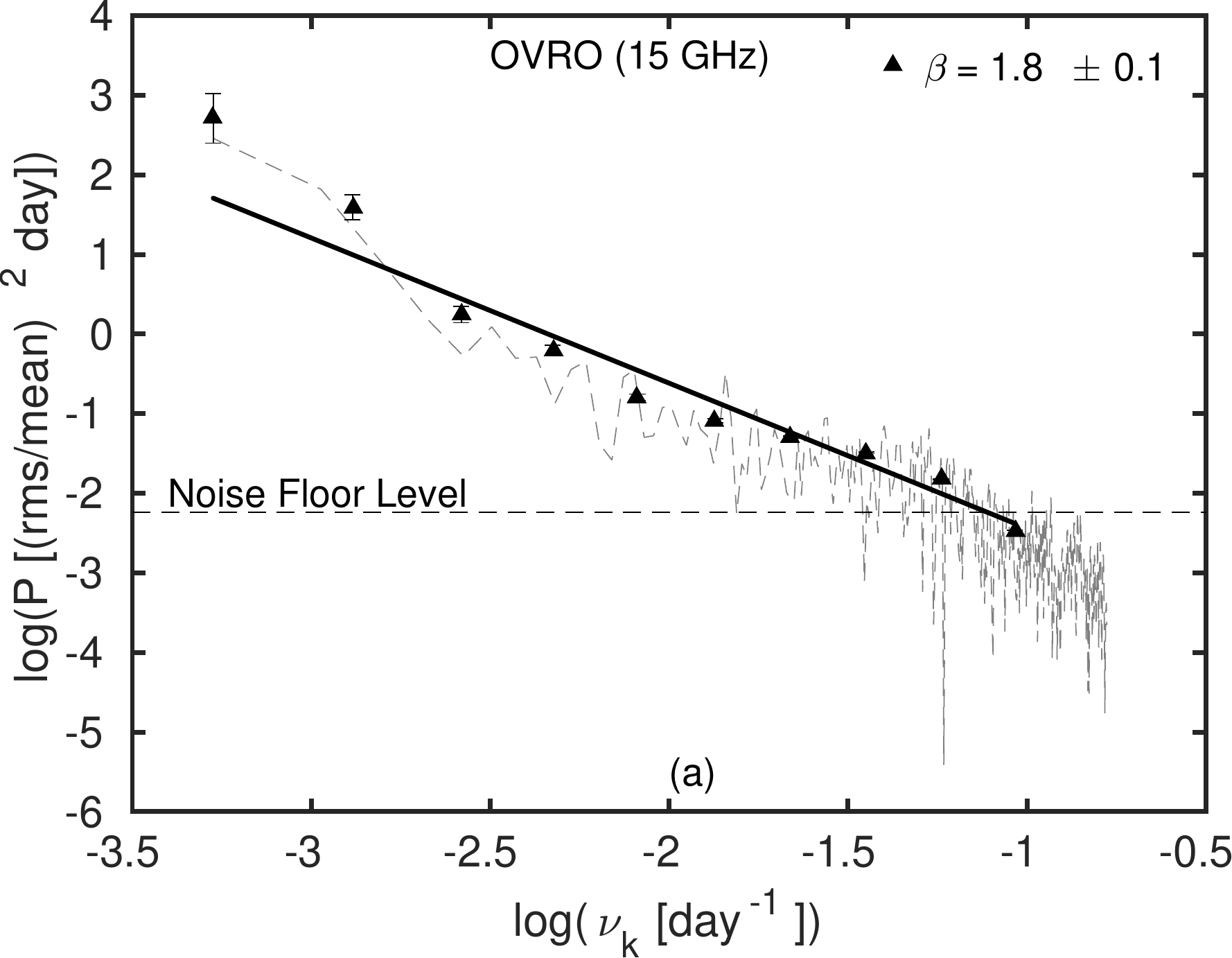}
\includegraphics[width=0.83\columnwidth]{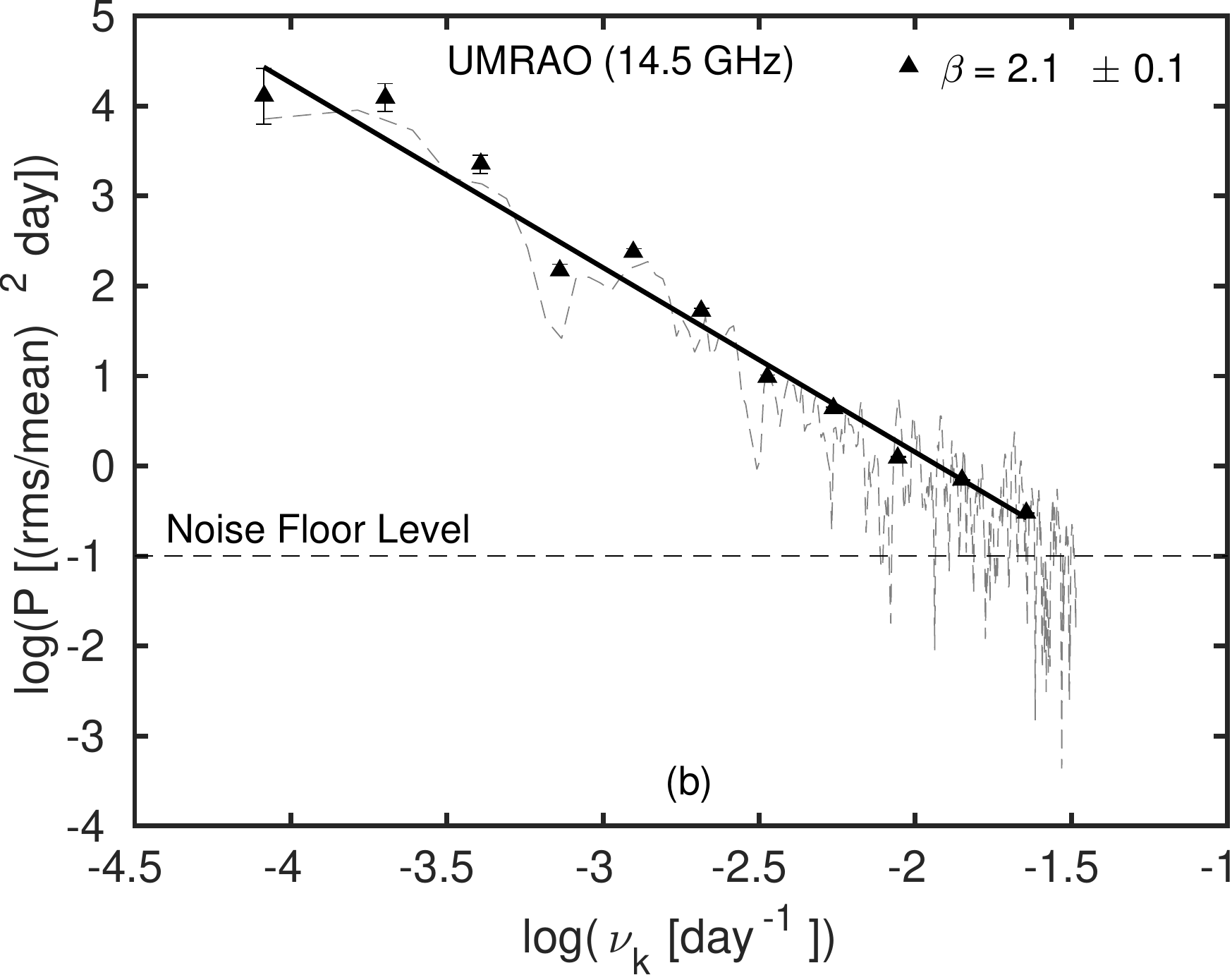}
\includegraphics[width=0.83\columnwidth]{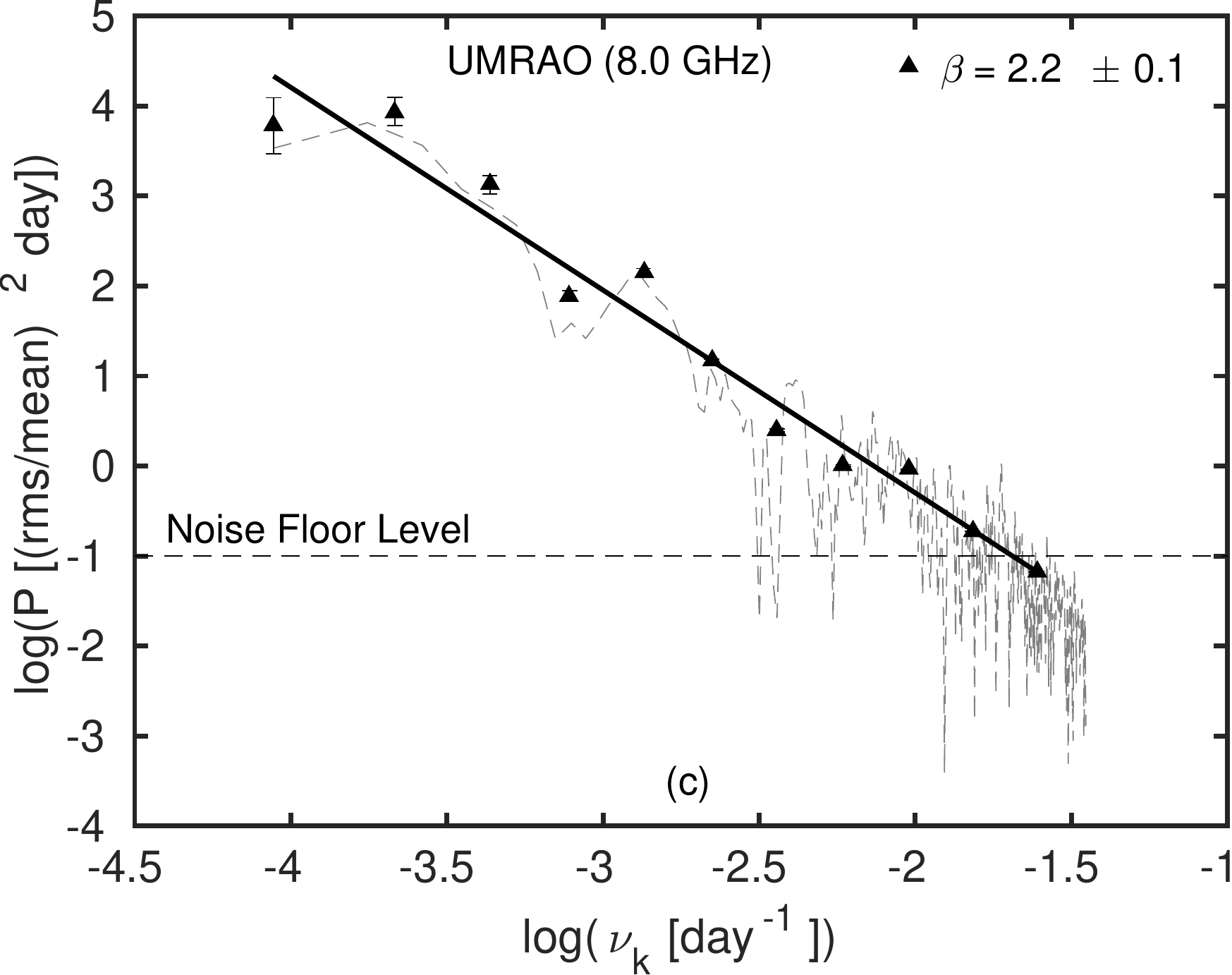}
\includegraphics[width=0.83\columnwidth]{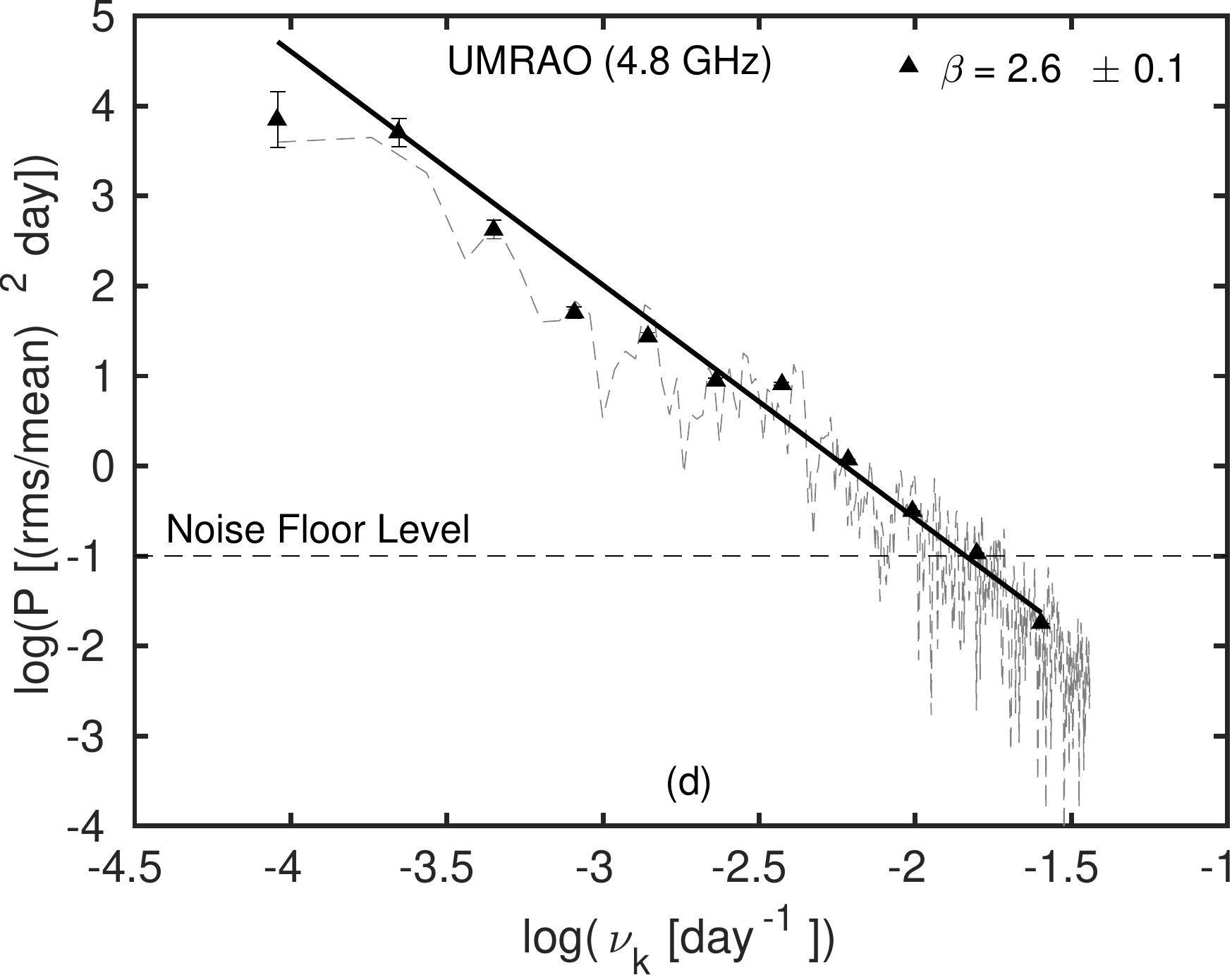}
\caption{As in Fig.\ 4 for the radio (OVRO and UMRAO) PSDs of PKS\,0735$+$178.}
\label{fig:radio}
\end{figure}

\begin{table*}[t!]
\caption{The PSD analysis for the light curves of PKS\,0735+178}
\label{tab:PSD}
\small
\centering
\begin{tabular}{ccccccccccc}\\\hline
Light curve & Range       &   $N_{obs}$       & $\Delta T_{\rm obs}$  &  $\Delta T_{\rm intep}$  & $T_{\rm obs}$   &$\rm \log(P_{stat})$ & $\log (\nu_k) $ range &  {$\beta \pm err$} \\     
            &           &        &      (day)     &   (day)             &            &             ((rms/mean)$^2$ day)       &  (day$^{-1})$      &     bin           \\
     (1)    &   (2)    &   (3)         &   (4)               &  (5)       &   (6)     &      (7)       &    (8)   & (9)           \\        
\hline
&&&&&&&&\\
Fermi-LAT (LT)        &  0.1-200 (GeV) & 168      &  15               &    0.5              &  6.9 (yr)    &  $+$0.50   &  $-$3.4 to $-$1.5           &1.1$\pm$0.2 \\
OVRO   (LT)           &   15  (GHz)     & 263      &  3                &    0.5              &  5.1 (yr)    &  $-$2.24   &  $-$3.3 to $-$1.0           &1.8$\pm$0.1 \\
UMRAO  (LT)           &  14.5 (GHz)     & 359      &  15               &    0.5              &  33  (yr)    &  $-$1.00   &  $-$4.1 to $-$1.6           &2.1$\pm$0.1 \\
UMRAO  (LT)           &  8.0  (GHz)     & 299      &  15               &    0.5              &  31  (yr)    &  $-$1.06   &  $-$4.1 to $-$1.6           &2.2$\pm$0.1 \\
UMRAO  (LT)           &  4.8  (GHz)     & 220      &  15               &    0.5              &  30  (yr)    &  $-$0.99   &  $-$4.0 to $-$1.6           &2.6$\pm$0.1 \\
Optical (LT)          &  R-band         & 677      &  1                &    0.5              &  22.6 (yr)   &  $-$2.27   &  $-$3.8 to $-$1.1           &2.0$\pm$0.1 \\
2004 Dec 10 (IN) &  R-band        & 30       &  0.008            &    0.0004           &  6.2 (hr)    &  $-$7.23   &  $+$0.6 to $+$1.8           &2.8$\pm$0.7 \\
2004 Dec 23 (IN) &  R-band        & 13       &  0.017            &    0.0004           &  5.9 (hr)    &  $-$6.83   &  $+$0.6 to $+$1.3           &4.1$\pm$0.3 \\
2005 Jan 5 (IN)   &  R-band        & 26       &  0.008            &    0.0004           &  5.2 (hr)    &  $-$7.23   &  $+$0.8 to $+$1.8           &2.2$\pm$0.3 \\
2005 Jan 9 (IN)   &  R-band        & 30       &  0.008            &    0.0004           &  7.1 (hr)    &  $-$7.21   &  $+$0.5 to $+$1.7          &1.5$\pm$0.2 \\
2007 Dec 15 (IN) &  R-band        & 29       &  0.008            &    0.0004           &  7.0 (hr)    &  $-$7.10   &  $+$0.5 to $+$1.7           &2.3$\pm$0.5 \\
2013 Jan 7 (IN)   &  R-band        & 22       &  0.008            &    0.0004           &  5.6 (hr)    &  $-$6.73   &  $+$0.5 to $+$1.6          &3.3$\pm$0.4 \\
\hline                                                                                        
\end{tabular}
\begin{minipage}{\textwidth}
Columns : (1) Light curve/date of the observation in the case of the optical intra-night datasets (LT: long-term, IN: intra-night); 
(2) the observed photon energy/frequency;
(3) the number of data points in the observed light curve;
(4) the typical sampling interval for the observed light curve; 
(5) the sampling interval for the interpolated light curve; 
(6) the total duration of the observed light curve (yr- year, hr- hour);
(7) the noise floor level in PSD due to the measurement uncertainty;
(8) the temporal frequency range covered by the   binned logarithmic power spectra;
(9) the power-law slope of the PSD along with the corresponding errors, for the binned logarithmic power spectra (see \S~\ref{sec:PSD}).    
\end{minipage}
\end{table*}

The periodogram obtained using equation~(\ref{eq:psdeq}), with the discrete set of frequencies $\nu_{k} = k/T$ with $k=1, ..., N/2$, is known as the `raw' periodogram. It consists of independently distributed $\chi^2$ variables with two DOF. This means that the dispersion in each estimate around the true value equals the true value itself, providing a noisy estimate of the spectral power \citep{1993MNRAS.261..612P, 2003MNRAS.345.1271V}.  To circumvent this problem, we have also constructed `binned' logarithmic periodograms, following the procedure outlined in \citet{1993MNRAS.261..612P}, \citet{2005A&A...431..391V}, and \citet{2015ApJ...798...27I}. We binned our periodograms with a constant factor of 1.6 in frequency (and by a factor of 1.5, in case of periodograms derived for the long-term optical light curve) and evaluated the mean power at the representative frequency taken as the geometric mean of each bin. The derived periodogram displays constant variance around the true underlying power spectrum, as if we are observing a noise process following the $\chi^2$ distribution with two DOF. For the binned logarithmic periodogram the variance decreases by a factor of $0.310/M$, where $M$ is  the number of data points in each bin. Furthermore, the true power spectrum is related to the observed power spectrum as $P(\nu_k) = (\chi^2/2) \, P_{true}({\nu_k})$. In order to derive the PSD slope we fit the power law function using a least-squares fit method in log-log space. Since the scatter is multiplicative in the linear space, it is additive in log-log space and identical at each frequency, so
\begin{equation}
 \log [ P(\nu_k) ] = \log [ (P_{true}({\nu_k}) ] + \log [ \frac{\chi^2}{2} ] .
\label{logpsd}
\end{equation}
Hence, the expectation value of the periodogram in log-log space is not the same as the expectation value of the power spectrum: there is a bias between the two values. This bias is a constant because of the shape of the $\chi^2$ distribution in log-log space. The expectation value of $\log [\chi^2/2] = -0.25068 $ \citep{1993MNRAS.261..612P}. Therefore, this value is added in the estimate of the binned periodogram. 

\begin{figure*}[t!]
\centering
\includegraphics[width=\textwidth]{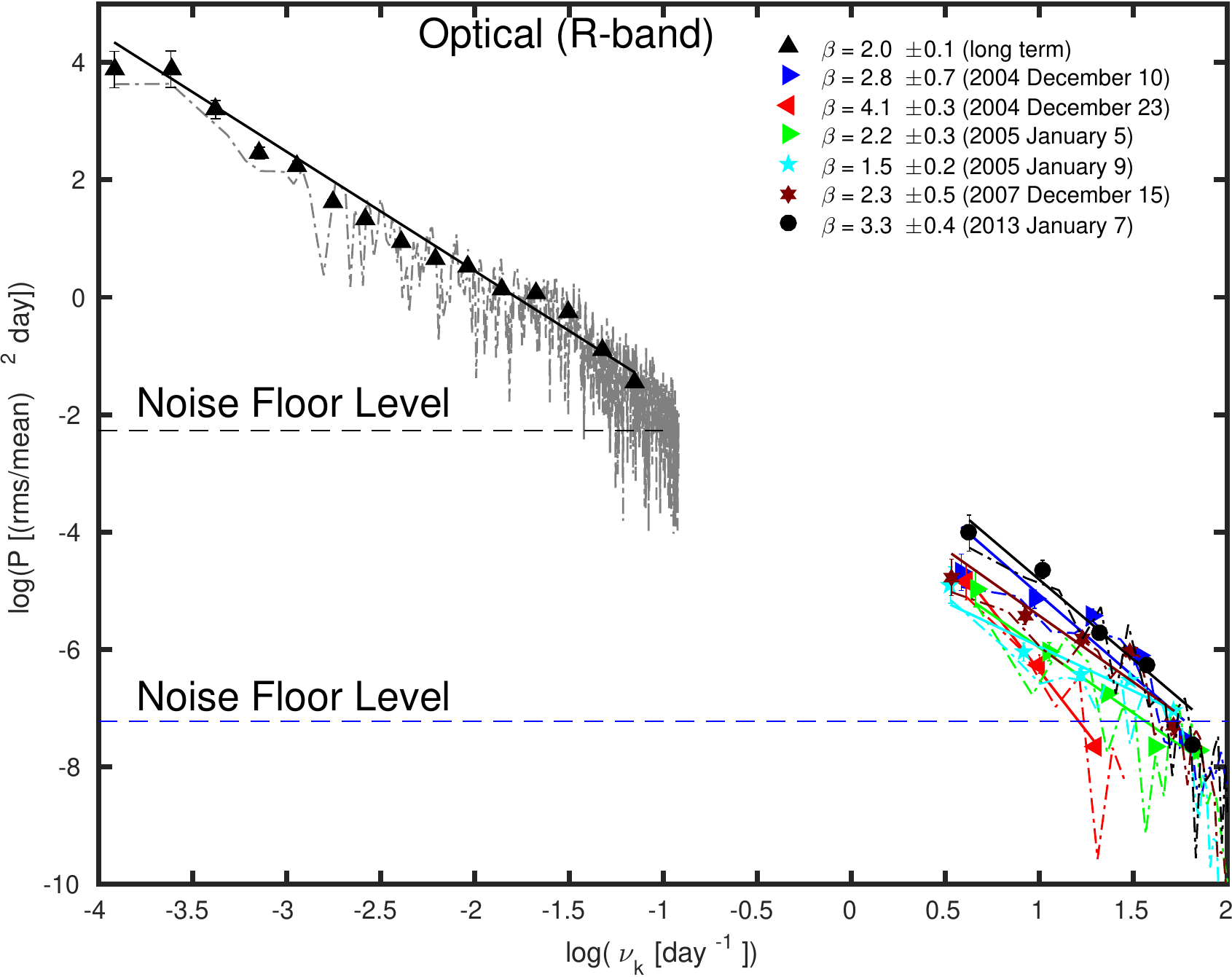}
\caption{Composite (long-term + intra-night) optical R-band PSD of PKS\,0735+178. Filled symbols denote the binned logarithmic periodogram estimates, respectively, for each data set considered (denoted here with long-term for the long-term optical monitoring data and the Julian date for the intra-night data). The gray dot-dashed lines shows the raw periodogram for the long-term optical data while the same color (as that of filled symbols) has been chosen to display the `raw' periodograms for the intra-night optical data.   The black dashed horizontal lines correspond to the noise floor level due to measurement uncertainty in long term optical data while blue dashed horizontal line show the typical noise floor level due to measurement uncertainty in intra-night data sets (see Table 2), as achieved in our observations.}
\label{fig:optical}
\end{figure*}

All the generated PSDs are summarized in Table~\ref{tab:PSD}, and presented in Figures~4 ({\it Fermi}-LAT), \ref{fig:radio} (UMRAO and OVRO), and \ref{fig:optical} (optical R-band) for the actual durations of the corresponding light curves, down to the observed sampling intervals. We have not subtracted the constant noise floor level (shown by the dashed horizontal lines in the figures), as some of the data points are below this level. For the analysis, the logarithmic binned PSDs were fitted with a single power law model $P(\nu_k)\propto{\nu_k^{-\beta}}$ using linear regression with weighted error in log-log space; the results of the fitting, along with the errors calculated as the rms residuals between the model and the data, as well as the corresponding variability frequency ranges of the analyzed light curves, are summarized in Table~\ref{tab:PSD}.

Figure\,\ref{fig:optical} presents a composite PSD for the optical data set analyzed in this paper, including the long-term monitoring data (1993--2015; 23 years down to nightly sampling), as well as the intra-night data with confirmed INV detections (spanning 4--8\,h down to $\sim 10-15$\,min sampling; see Table~\ref{sec:obs}). The fitted power-law slope is $2.0\pm0.1$ for the long-term segment of the optical PSD, and ranges from $\sim 1.5$ up to $\sim 4.0$ for the individual intra-night data sets. The composite optical PSD (long-term + intra-night) covers an unprecedented frequency range of nearly 6 dex for a blazar source. This range, from, $\sim 10^{-9} - 10^{-3}$\,Hz, is primarily due to the high photometric accuracy achieved in our intra-night monitoring program. In all the cases, however, the normalization of intra-night PSDs turns out to be consistent with a simple extrapolation of the red-noise ($\beta \sim 2$) optical PSD from lower temporal frequencies.

Figure\,\ref{fig:psd_mf} presents the composite multi-wavelength PSDs of PKS\,0735+178 corresponding to the variability timescales $>10$\,days.  This explicitly shows the similarities between the radio and optical bands and the clear difference of the $\gamma$-ray band from the others for which we have enough data to compute sensible PSDs.

\section{Discussion and Conclusions}

The main findings from our analysis of the multi-wavelength, multi-epoch, photo-polarimetric data for PKS\,0735+178, can be summarized as follows:
\begin{enumerate}
\item{ 
The INV duty cycle of PKS\,0735$+$178 in the extended dataset consisting of 25 nights (22 from our monitoring program and three from the literature; Figure~\ref{fig:INV}), over the time span of 18 years, is DC\,$\sim 22\%$ for all the variability amplitudes $\psi$, but DC\,$\sim 4\%$ when only the nights showing $\psi > 3\%$ are considered. This value of INV DC is $\sim$ 5 --10 times smaller than that typically observed in blazar sources, indicating that PKS\,0735$+$178 remains in a state of optical quiescence on intra-night timescales, in conformity with our previous study \citep{2009MNRAS.399.1622AG}.  
}
\item{
The newly obtained long-term {\it Fermi}-LAT and optical data reveal weakly-correlated, a-factor-of-a-few, flux changes in the source on month-like timescales (Figure~\ref{fig:ZLC}); meanwhile, the archival radio and optical data show uncorrelated, larger-amplitude (factor of several) variability over the timescale of years (Figure~\ref{fig:LC}). 
}
\item{
The gathered photo-polarimetric data reveal large variations in the optical PA, traced during the period 2005--2015 on timescale of months to years; those changes seem erratic, with no clear repetitive pattern and no correlation with either the total optical flux, the optical PD, or even with the overall level of the optical polarization degree ($\lessgtr 3\%$; Figure~\ref{fig:ZLC}). Interestingly, the optical PD of this blazar has frequently been found to be below 3\%.
}
\item{
The PSDs of PKS\,0735$+$178 at radio (GHz) and optical frequencies, on the timescales from years to weeks, resemble each other closely: both can be well represented by a single power-law function with the slope $\beta \simeq 2$, meaning a ``pure red noise'' type of the source variability at the lower frequencies of the electromagnetic spectrum. On the other hand, the high-energy $\gamma$-ray PSD, within the corresponding temporal frequency range, is best fitted with a much flatter slope of $\beta\simeq 1$, corresponding to a ``pink noise'' behavior (Figure\,\ref{fig:psd_mf}); no low-frequency flattenings in the PSDs have been detected.
}
\item{
The optical PSDs on intra-night timescales (for the nights during which INV has been detected), are characterized by a range of slopes mostly between 1.5 and 3.3 with the exception of one instance where it was $\sim$4 (however, this was from a shorter observation with relatively long integration times so that it was derived using only 3 points in the binned periodogram).} This implies a non-stationarity of the variability process in the source, at least on hourly timescales; in all the cases, however, the normalized intra-night PSDs, is consistent with a simple extrapolation from the red-noise ($\beta \sim 2$) optical PSD observed on longer timescales (Figure~\ref{fig:optical}, Table~\ref{tab:PSD}). No high-frequencycut-offs have been detected in the intra-night power spectra down to the noise floor levels.
\end{enumerate}

\subsection{Polarization and microvariability}

The very low INV DC of PKS\,0735+178 reported previously in the literature \citep{2009MNRAS.399.1622AG} was puzzling, in view of the persistently high optical polarization degree claimed for the source \citep{1996AJ....112.1877G, 2011ApJS..194...19W}. Our extended monitoring has confirmed the low INV DC but, at the same time, revealed that the ${\rm PD_{opt}}$ of the blazar frequently drops below $3\%$  -- for about one-third of the monitoring time, based on the available optical polarimetry. 

Traditionally, the degree of optical polarization has been used to differentiate between blazars and LPQs \citep{1980ARA&A..18..321A, 1984ApJ...279..465M}. In blazar sources, states with very high $\rm PD_{opt}$ reaching even $\sim 40-50\%$, are not uncommon \citep[e.g.,][]{2002PASJ...54L..55F, 2013ApJ...768...40L}, indicating highly ordered magnetic fields within the jet regions where the observed optical synchrotron emission is being produced; in the case of LPQs, on the other hand, the optical polarization is presumed to be reduced predominantly through the more significant contribution of (unpolarized) thermal emission from an accretion disk to the radiative output of a source at optical wavelengths \citep[e.g.,][]{1997A&A...325..109S}. Yet, PKS\,0735+178 is a typical example of a BL Lac object lacking any strong emission lines, while those would be expected to be visible in the presence of pronounced disk continuum emission \citep{2012A&A...547A...1Nilsson}.

Although some blazars are known to alter between the ``blazar-like'' and the ``LPQ-like'' optical polarization levels \citep[e.g.,][]{1988A&A...205...86F}, the frequency of such transitions is poorly known, presumably due to the lack of regular polarimetric observations. Only a few of the brightest sources have been subjected to long-term optical photo-polarimetric monitoring.  For example, the BL Lac object OJ\,287 during the period 2005--2009, displayed $\rm PD_{opt} < 3\%$ for only $\lesssim 0.5\%$ of the total observed time span \citep{2010MNRAS.402.2087V}. Hence, one may conclude that PKS\,0735$+$178 is a peculiar blazar in showing \emph{both} a relatively low microvariability duty cycle, \emph{and} frequent states of low optical polarization.

Based on a comparative study of quasars of the HPQ and LPQ types, \citet{2012A&A...544A..37AG} concluded that the pronounced and frequent INV is correlated with the overall high degree of the optical polarization, rather than with the degree of relativistic beaming, which seems to play a secondary role. Specifically, blazars and quasars with similar degrees of relativistic beaming \citep[i.e., with a similar radio core prominence; e.g.,][]{1982MNRAS.200.1067Orr}, exhibit  considerably and systematically different INV DCs. This is in agreement with our new results summarized above for PKS\,0735$+$178, suggesting a basic link between the magnetic field ordering in the jet, as reflected in the optical polarization degree, and the production of rapid flux changes. 

We finally note in this context that the VLBI radio jets of HPQs show $\chi_{\rm rad}$ of the core/inner jet region parallel to the jet direction, and typically well aligned with $\chi_{\rm opt}$, while the VLBI radio jets of LPQs are characterized by $\chi_{\rm rad}$ misaligned with respect to the jet axis, and uncorrelated with $\chi_{\rm opt}$ \citep{2000ApJ...541...66L}. For PKS\,0735$+$178, we observe erratic changes of $\chi_{\rm opt}$ between 0 and 180\,deg, uncorrelated with $\chi_{rad}$ and with no special relation to the VLBI jet direction in the source \citep[the VLBI jet position angle is $\sim 70$\,deg;][]{2010A&A...515A.105B}. Such a behavior is hard to reconcile with a ``grand design'' helical magnetic field structure inferred for the source by \citet{2006MNRAS.369.1596G} and \citet{2006A&A...453..477A}. 

\subsection{Power spectra}

The LT radio PSDs of blazar sources analyzed in the literature within the temporal frequency range $\sim 10^{-9} - 10^{-7}$\,Hz, are best represented by single power-laws with slopes $\beta \sim 1.5-2.5$  \citep{2014ApJ...785...76P, 2014MNRAS.445..437M}. Specifically, the PSD of the BL Lac object PKS\,2155$-$304, while consistent with a power-law function $\beta \sim 1.8$, shows a  flattening at frequencies below $\sim 10^{-3}$\,d$^{-1}$, suggesting a transition from a red noise to a white noise type of variability on the timescale of a few years \citep{2011A&A...531A.123K}. For PKS\,0735$+$178, \citet{2007A&A...467..465C}  analyzed the long-term optical monitoring data (1970--2004) using the structure function method and found the associated PSD slope $\beta$ to be between 1.5 and 2.0 on the timescales ranging from 33 years down to weeks; at the same time, they found a variety of PSD slopes, $\beta  \sim 1.5-2.3$, for densely-sampled light curves from separate observing seasons (covering the timescales of a few months to a few days). Also, the optical PSDs of blazars detected with the Kepler telescope are characterized by $\beta \sim1.5-2.0$ within the temporal frequency range $\sim 10^{-7} - 10^{-4}$ Hz \citep{2014ApJ...785...60R, 2013ApJ...766...16E}. 

The X-ray PSDs of BL Lacs, on the other hand, are typically consistent with a broken power-law model with the slopes $\beta \sim 2-3$ above the break frequency $\sim 1$\,d$^{-1}$, and $\beta \sim 1-2$ below the break \citep{1985ApJ...296...46S, 2001ApJ...560..659K,2002ApJ...572..762Z,2003A&A...402..929B, 2013ApJ...770...60S,2015ApJ...798...27I}.  Unfortunately, PKS\,0735$+$178 is quite weak in X-rays, so very few observations of it have been made and it is not possible to compute a sensible X-ray PSD for it.  The {\it Fermi}-LAT PSD slopes for the bright BL Lacs and FSRQs, have been estimated by \citet{2010ApJ...722..520A} as $\beta \sim 1.7 \pm 0.3$ and $\sim 1.4 \pm 0.3$, respectively, for  temporal frequencies from $\sim 10^{-8}$ to $10^{-6}$\,Hz \citep[see also in this context][]{2013ApJ...773..177N,2010ApJ...721.1383A}.  \citet{2014ApJ...786..143S}, who modeled in detail the 4\,yr-long {\it Fermi}-LAT light curves of the brightest blazar sources down to week-long sampling intervals, claimed the corresponding PSD slopes to be typically $\beta \simeq 1$, in a very good agreement with our result for PKS\,0735$+$178.

Rather limited work has been carried out to compare the PSD properties at different wavelengths of the electromagnetic spectrum for a given blazar source. Most notably, \citet{2008ApJ...689...79C}, by comparing the LT PSD of the luminous FSRQ 3C\,279 at X-ray (3--20\,keV), optical (R-band), and radio (GHz) frequencies, covering the total time span of $\sim 11$ years (temporal frequencies from $10^{-8.5}$\,Hz down to $10^{-5}$\,Hz), demonstrated that each PSD can be well fitted by a single power-law with $\beta \sim 2.3$ (X-rays),  $\sim 1.7$ (optical), and $\sim 2.3$ (radio). A more extensive study by \citet{2012ApJ...749..191C}, consisting of six blazars observed with {\it Fermi}-LAT and optical/near infra-red telescopes, albeit only for a total duration $\sim 1$ year (temporal frequency range $\sim 10^{-7.5} - 10^{-5.5}$\,Hz), revealed that, on average, the blazar PSDs at different frequencies are roughly consistent with  $\beta \sim 1.6$.

Only a few blazar studies have addressed the issue of the PSD characteristics on intra-night timescales. \citet{2003A&A...397..565P} estimated the optical PSD slopes for BL Lacertae to be $\beta \simeq 1.87 \pm 0.16$ on five observing nights within the frequency range $\sim 10^{-4} - 10^{-3}$\,Hz. In the case of a highly variable BL Lac object 0716+714, \citet{2005IAPPP.101....1A} noted rather flat optical PSD slopes with $\beta \sim 0.9-1.3$ on 10 monitoring nights for the frequency range $\sim 10^{-5} - 10^{-3}$\,Hz; meanwhile, \citet{2011AJ....141...49C} estimated $\beta \sim 2-3$ on five nights and within the same temporal frequency range \citep[see also][]{2012Ap&SS.342..147M}. The wide range of PSD slopes cited above on intra-night timescales for 0716$+$716,  implies either some significant inconsistency between the different analysis methods applied, or the non-stationarity of blazar variability on hourly timescales. The latter case echoes our results for PKS 0735$+$178.

Finally, in the very high energy $\gamma$-ray domain, \citet{2007ApJ...664L..71A} reported the PSD slope of $\beta \sim 2$ for $\nu_k \sim 10^{-4} - 10^{-2}$\,Hz during the famous TeV outburst of PKS\,2155$-$304.  

\begin{figure}[t!]
\centering
\includegraphics[width=\columnwidth]{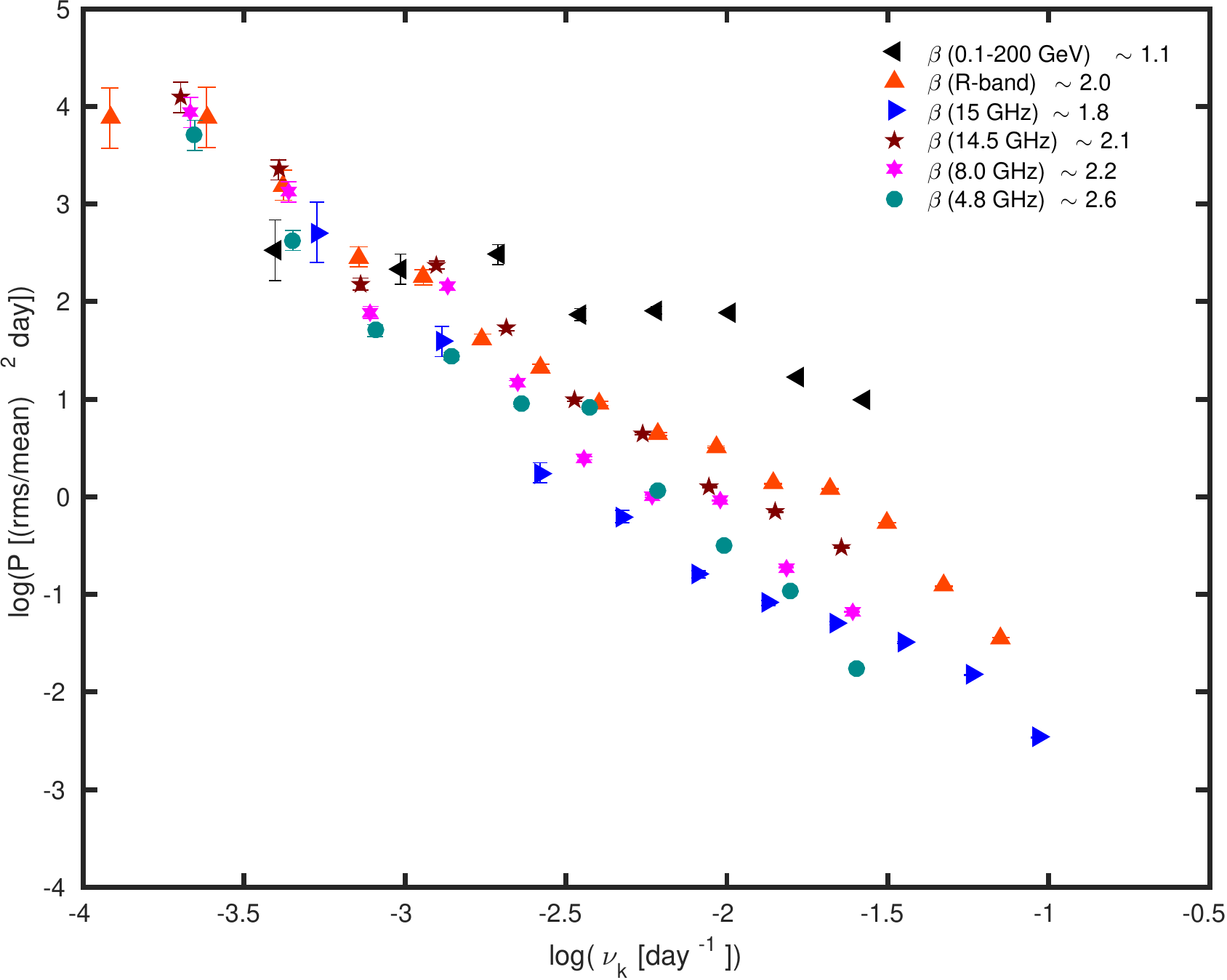}
\caption{Composite multiwavelength PSD of PKS\,0735+178. Symbols denote the binned logarithmic periodogram estimates for each data set considered.}
\label{fig:psd_mf}
\end{figure}

The PSD slopes derived from our detailed analysis of the LT high energy $\gamma$-ray (0.1--200\,GeV), optical (R-band), and radio (GHz frequencies) data on PKS\,0735$+$178, on the timescales ranging from years to weeks/days, suggest that the statistical character of the $\gamma$-ray flux changes is different from that of the radio and optical flux changes: there is increasingly more variability power in $\gamma$-ray fluctuations on comparable temporal frequencies when going to shorter and shorter variability timescales (see Fig.\ \ref{fig:psd_mf}). Our finding is in agreement with the {\it Fermi}-LAT blazar data analysis presented by \citet{2014ApJ...786..143S}. This result is somewhat surprising since, at least within the framework of the standard one-zone leptonic models for the broad-band blazar emission, which is supported to some extent by the multi-wavelength correlations sometimes observed \citep[e.g.,][]{2014MNRAS.439..690H, 2012A&A...537A..32A}, the power spectrum of the synchrotron emission component (radio and optical photon energies) should be the same as, or eventually flatter than, the power spectrum of the inverse-Compton component \citep{2014ApJ...791...21F}.

As discussed by \citet{2009ApJ...698..895K}, a broken power-law form of the PSD with flat low-frequency segment $P(\nu_k) \propto const$ and the high-frequency slope $\beta = 2$, lacking any peaks indicative of a (quasi-)periodicity, can be understood in terms of a source variability being driven by an underlying stochastic process. In particular, \citet{2009ApJ...698..895K} proposed that such a variability can be modelled as a first-order autoregressive process (Gaussian Ornstein-Uhlenbeck process; OU for short), in which the source emissivity responds to some input noise (Gaussian white noise, by assumption), with a given relaxation timescale $\tau_1$. For such a situation, the source variability at temporal frequencies above the break $\nu_{k_1} \equiv (2 \pi \tau_1)^{-1}$ is of the red noise type, and below the break, of the white noise type. 

\citet{2011ApJ...730...52K} discussed a more complex case of a linear superposition of OU processes with two very different relaxation timescales, $\tau_1 > \tau_2$, resulting in the intermediate pink noise ($\beta = 1$) segment in the source PSD, in between of the white noise ($\beta=0$) below $\nu_{k_1}$ and the red noise ($\beta = 2$) above $\nu_{k_2}$.

The shallower slopes of the high-energy $\gamma$-ray PSD (as compared to the optical and radio PSDs) could be therefore understood by hypothesizing, following \citet{2011ApJ...730...52K}, that the $\gamma$-ray variability in PKS\,0735$+$178 is shaped by a linear superposition of  two types of stochastic processes: (i) the same process driving synchrotron (optical and radio) variability, with the relaxation time $\tau_1$ larger than decades (since we do not recover the transition from the red noise to the white noise in the optical PSD up to  timescales of $\sim 20$ years); and (ii) the additional process relevant only in the inverse-Compton ($\gamma$-ray) domain, characterized by a relaxation timescale $\tau_2$ shorter than, or at most of the order of, days (the shortest variability timescales covered by our analysis of the {\it Fermi}-LAT PSD).

Note that in this scenario there need not be some instability operating within the jet on the timescale $\tau_1$ which is driving the synchrotron variability -- the ``drivers'' can instead be random (stochastic) fluctuations in local jet conditions. Such fluctuations dissipate their energy and accelerate plasma particles, thus creating fluctuations in the distribution of ultra-relativistic jet electrons. The acceleration process itself and the radiative response of the accelerated electrons are, however, delayed with respect to the input perturbations by $\tau_1$, so that in the time domain below $\tau_1$ the flux changes are smoothed out and damped (forming the red noise segment of the PSD). The jet adjusts, forgetting about the input perturbations, only on timescales longer than $\tau_1$, for which the flux changes become uncorrelated (forming the white noise segment of the PSD).

At this point we can only speculate about the nature of the input perturbations inferred above.  They could be identified with, for example, local fluctuations in the plasma bulk velocity, leading to the formation of a complex system of internal shocks and plasma turbulence, possibly passing through and interacting with a stationary reconfinement shock further down the jet \citep{2014MNRAS.443..299M,2014ApJ...780...87M,2015JApA...36..255C,2016ApJ...820...12P}. Alternatively, input disturbances could be related to a turbulent magnetic field inherited from strongly magnetized accretion disk and carried out by the jet (see in this context \citealt{2003ApJ...592.1042I,2012MNRAS.423.3083M}; also \citealt{2013ApJ...772...83L}). In both cases, a relatively long relaxation timescale $\tau_1$ would correspond to some global (magneto-)hydrodynamical timescale characterizing an extended blazar emission region. 

Meanwhile, in the $\gamma$-ray regime, the additional process should be related to inhomogeneities in the local (radiating fluid rest frame) energy densities of soft photons, inverse-Comptonized to higher energies by the jet electrons in the neighboring cells of the outflow. A natural ``relaxation'' timescale in this case should be simply a light crossing timescale for the jet region where the bulk of the observed emission is being produced. For $\tau_2 \lesssim $\,day, one  then has the corresponding jet radii $R_2 \sim c \tau_2 \, \langle \delta_{\rm j} \rangle \lesssim 0.03$\,pc, and the distances from the core $r_2 \sim R_2 \, \langle \Gamma_{\rm j} \rangle \leq 1$\,pc, where, for order-of-magnitude estimates only, we assumed a conical jet with a half-opening angle $\theta_j \simeq 1/\langle \Gamma_{\rm j} \rangle$, and the volume-averaged jet bulk Lorentz and Doppler factors, $\langle \Gamma_{\rm j} \rangle \simeq \langle \delta_{\rm j} \rangle \simeq 30$. 

In the interpretation mentioned above, which we will  discuss in detail in future work, we must revisit the overly simplistic assumption of a single homogeneous emission zone in blazar jets. Instead, we propose that \emph{all} the observed blazar emission, variable on  timescales of years, months, days, and hours --- i.e., both the long-term large-amplitude variability and the microvariability --- is generated by an underlying single {\it stochastic} process (radio and optical frequencies), or a linear superposition of such processes ($\gamma$-ray regime), within a highly non-uniform portion of the outflow extending from the jet base up to the $\lesssim$\,pc-scale distances.

\begin{acknowledgements} 
Authors wish to thank the referee for making several constructive comments on the manuscript. AG thanks Bindu Rani, Alok C.\ Gupta, Rumen Bachev, Margo Aller, Hugh Aller, Paul Smith and Svetlana Jorstad for kindly providing  data in electronic form. AG, {\L}S and MO acknowledge support from the Polish National Science Centre (NCN) through the grant 2012/04/A/ST9/00083. AG also acknowledges partial support from 2013/09/B/ST9/00026 and MS acknowledges the support of 2012/07/B/ST9/04404. VL acknowledges the support of Russian RFBR grant 15-02-00949 and St.\ Petersburg University research grant 6.38.335.2015. IA acknowledges support by a Ram\'on y Cajal grant of the Ministerio de Econom\'ia y Competitividad (MINECO) of Spain. Acquisition and reduction of the MAPCAT data was supported in part by MINECO through grants AYA2010-14844, AYA2013-40825-P, and AYA2016-80889-P, and by the Regional Government of Andaluc\'ia through grant P09-FQM-4784. The MAPCAT observations were carried out at the German-Spanish Calar Alto Observatory, which is jointly operated by the Max-Plank-Institut f\"ur Astronomie and the Instituto de Astrof\'isica de Andaluc\'ia-CSIC. PJW is grateful for hospitality at KIPAC, Stanford University, during a sabbatical visit. This research has made use of data from the University of Michigan Radio Astronomy Observatory which has been supported by the University of Michigan and by a series of grants from the National Science Foundation, most recently AST-0607523, and NASA Fermi grants NNX09AU16G, NNX10AP16G, and NNX11AO13G.
The OVRO 40m Telescope Fermi Blazar Monitoring Program is supported by NASA under awards NNX08AW31G and NNX11A043G, and by the NSF under awards AST-0808050 and AST-1109911.  Data from the Steward Observatory spectropolarimetric monitoring project were used. This program is supported by Fermi Guest Investigator grants NNX08AW56G, NNX09AU10G, NNX12AO93G, and NNX15AU81G.  
The paper uses optical photometric and polarimetric data from the BU blazar monitoring programme. The research at BU was supported in part by NASA grants NNX14AQ58G and NNX15AR34G. 
The Fermi-LAT Collaboration acknowledges support from a number of agencies and institutes for both development and the operation of the LAT as well as scientific data analysis. These include NASA and DOE in the United States, CEA/Irfu and IN2P3/CNRS in France, ASI and INFN in Italy, MEXT, KEK, and JAXA in Japan, and the K. A. Wallenberg Foundation, the Swedish Research Council, and the National Space Board in Sweden. Additional support from INAF in Italy and CNES in France for science analysis during the operations phase is also gratefully acknowledged.

\end{acknowledgements} 

\appendix

\section{Interpolation of unevenly sampled light curves}
\label{sec:App}

Here we discuss in more detail the problem of interpolation of an unevenly sampled light curve of a red noise-type astrophysical source, presenting in particular the results of our simulations using various techniques for estimating source power spectra, and various window functions. We conclude that, even though the distortion of the signal by the interpolation procedure is inevitable, it may be nonetheless properly recognized, while on the other hand no choice of a particular window function in unevenly sampled data can substitute for linear interpolation. That is to say, power spectral density derived for unevenly sampled colored-noise data \emph{without} linear interpolation, does not resemble the real periodogram regardless on a choice of a window function \citep[see in this context][]{D75,2002MNRAS.332..231U,2014MNRAS.445..437M}.

The model power spectrum \emph{derived} using Fourier decomposition of a discrete signal sampled $N$ times at fixed $\Delta t$ time intervals, and observed for a finite amount of time $T = N \times \Delta t$, is a convolution of a \emph{true} power spectrum and a spectral window. In the finite time series, for which the end points of a light curve are not matched (i.e., when the number of periods in the acquisition is not an integer in the frequency domain), variability power ``leaks'' in the periodogram from low to high frequencies, and this effect is known as a ``red-noise leakage''. This leakage can be minimized by multiplying the time series with a proper window function, which reduces the amplitude of the discontinuities at the boundaries of a given time series \citep{1992nrca.book.....Press}. On the other hand, due to the discrete sampling of a light curve, for each variability frequency $\nu$ an additional power is added in the derived periodogram at variability frequencies $2 \nu_{\rm Nyq}\pm \nu$, $4 \nu_{\rm Nyq}\pm \nu$, ..., where $\nu_{\rm Nyq}= (2 \, \Delta t)^{-1}$ is the Nyquist frequency. The net effect of such ``aliasing'' is known {\it a priori} for evenly sampled data, while it has to be investigated {\it a posteriori} (in the Fourier domain of the spectral window) in the case of uneven sampling \citep{D75}.

In our test simulations an artificial light curve is generated using the method of \citet{TK95}, assuming a pure red-noise ($\beta= 2$) power spectrum; the data points are evenly sampled with a sampling period of one day, and the total duration of the simulated time series is 1,000 days. The power spectrum is obtained using the discrete Fourier transform (DFT) with three various window functions applied --- rectangular, Bartlett (triangular), and Hanning window functions --- and later also using the Lomb-Scargle periodogram (LSP) method and the Fourier transform (FT) of the autocorrelation function (ACF).

\begin{figure}[t!]
\centering
\hspace*{0.5cm} \includegraphics[width=0.35\textwidth]{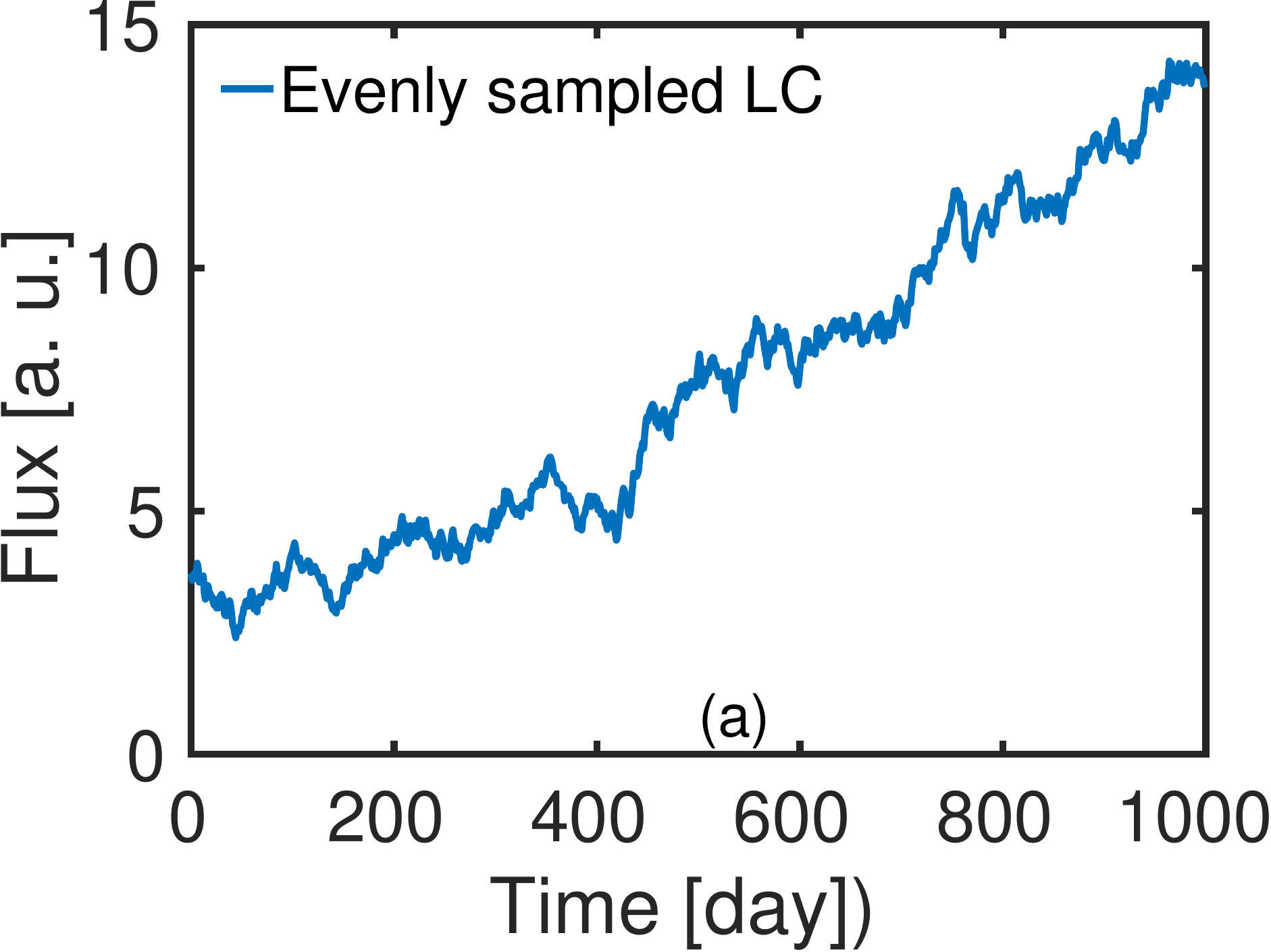} 
\includegraphics[width=0.35\textwidth]{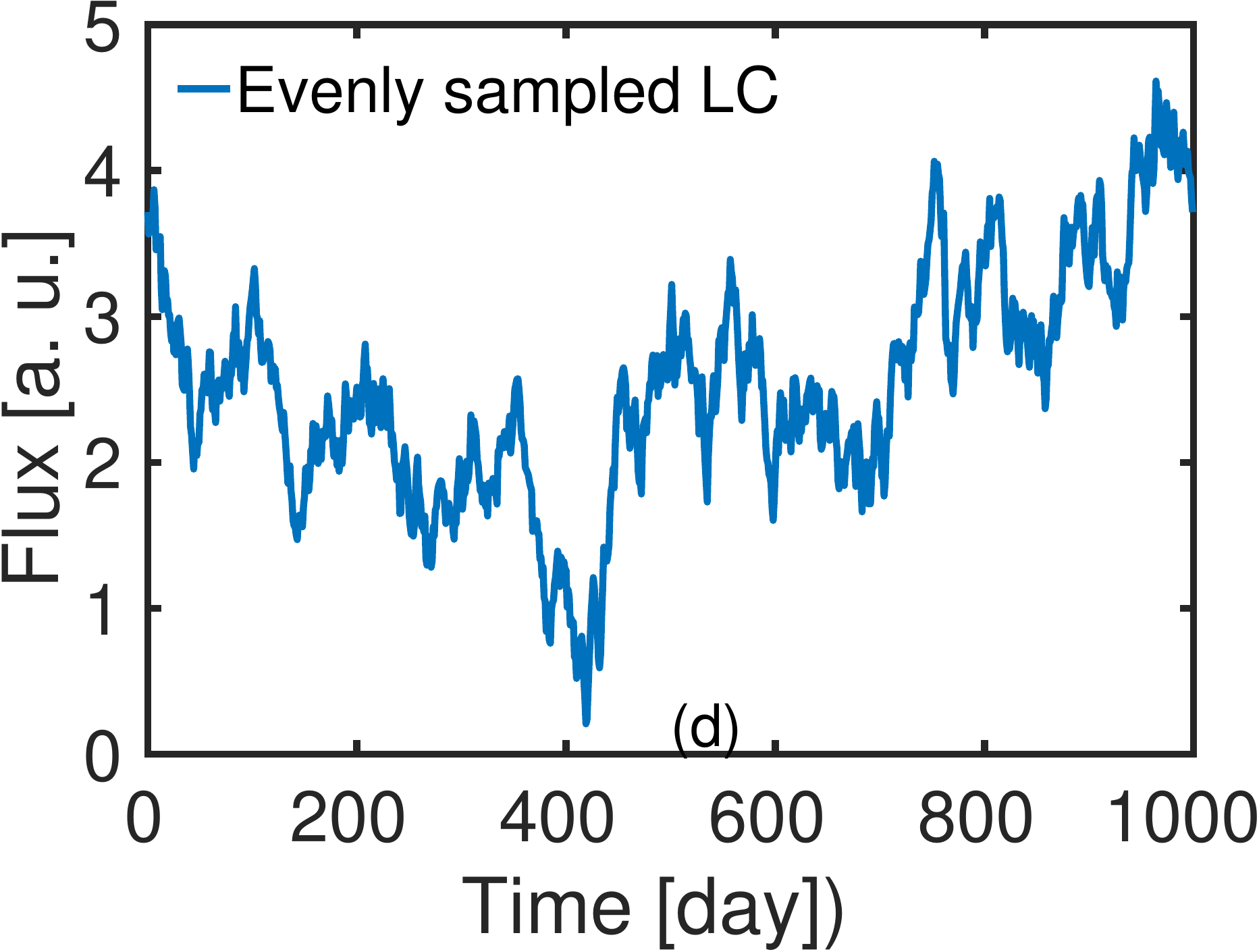} 
\includegraphics[width=0.35\textwidth]{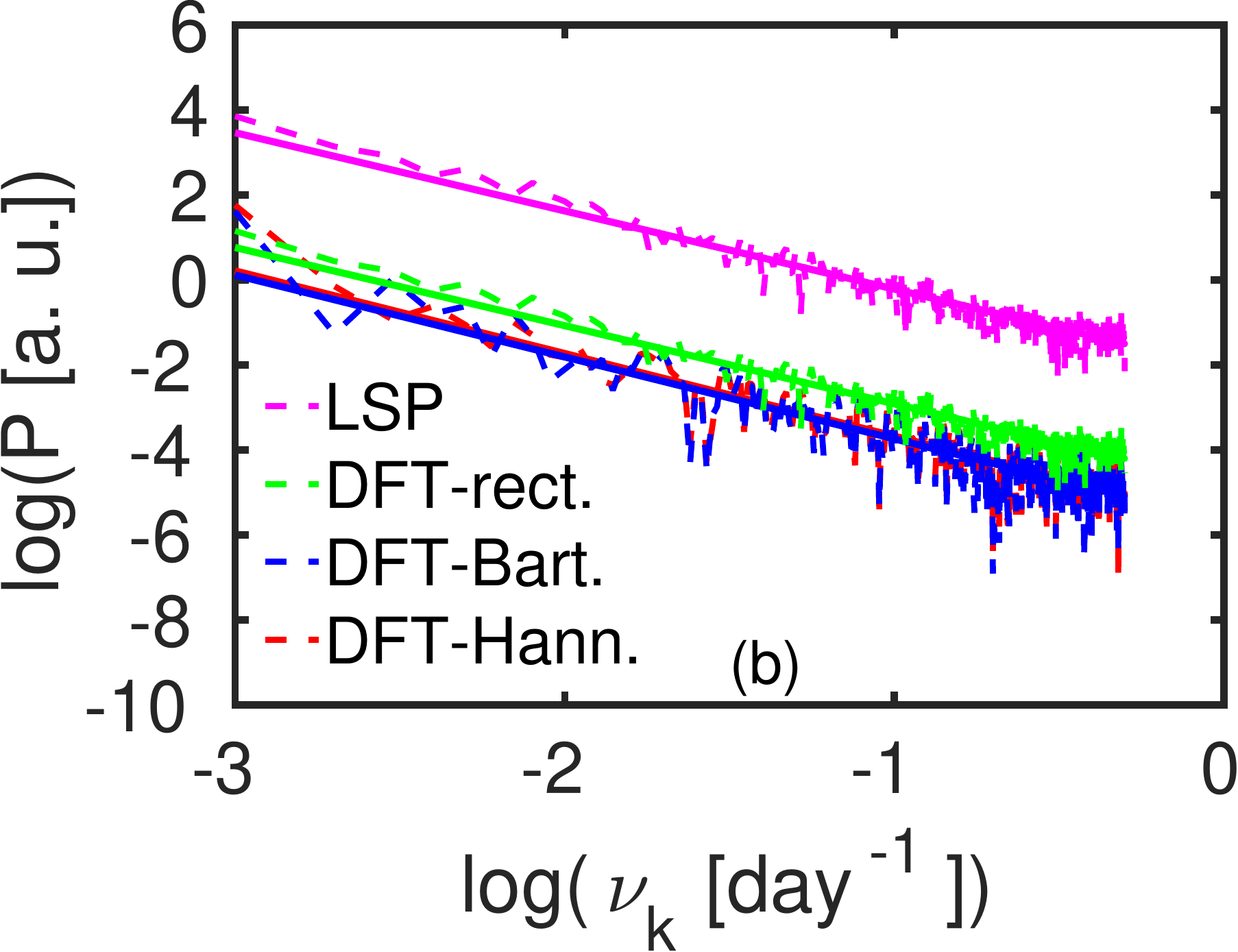}
\includegraphics[width=0.35\textwidth]{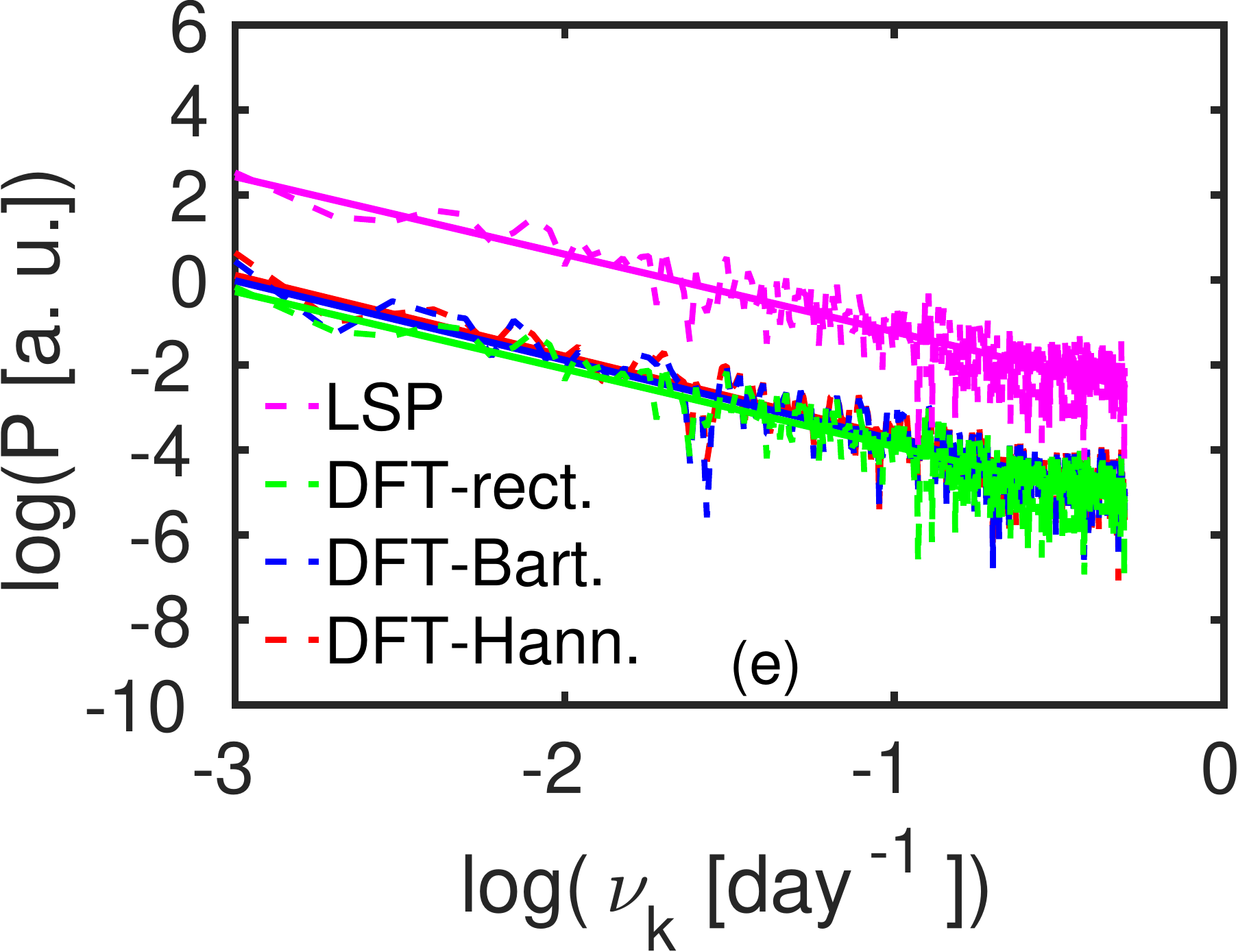}
\includegraphics[width=0.35\textwidth]{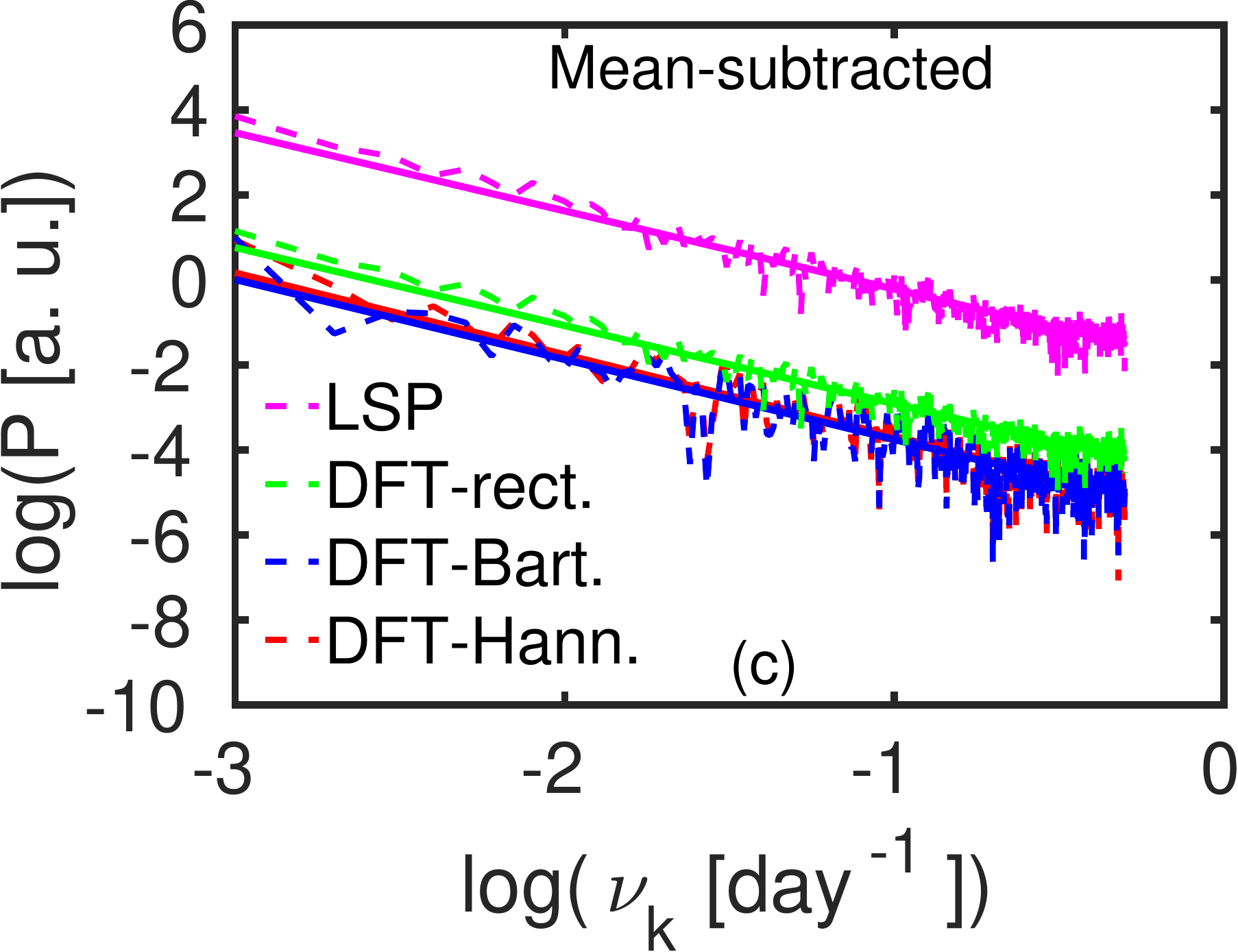}
\includegraphics[width=0.35\textwidth]{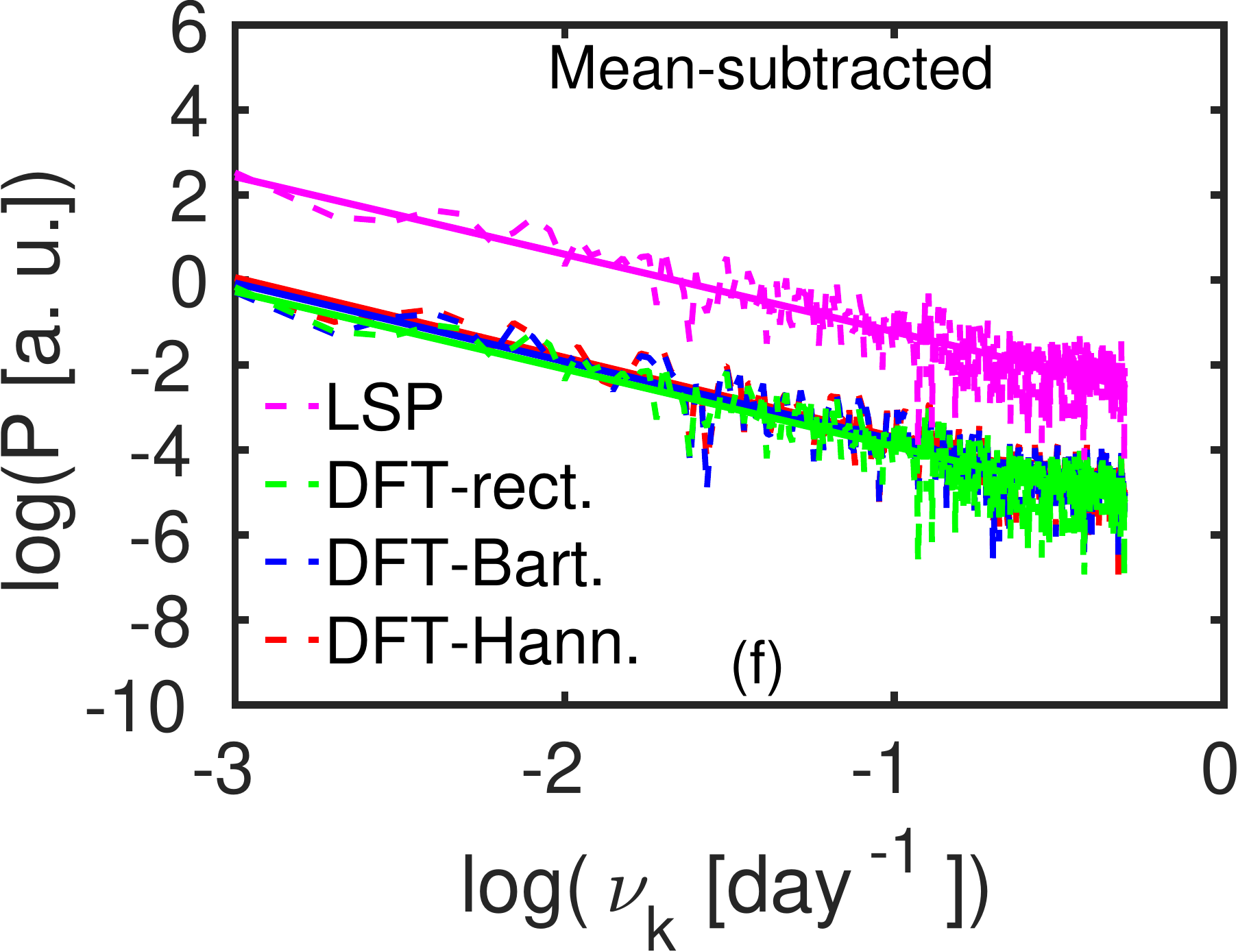}
\caption{{\it Top panels:} The simulated red-noise ($\beta=2$), evenly-sampled light curves. {\it Middle panels:} PSDs derived for the original simulated light curves using various methods, as denoted in the plots. {\it Bottom panels:} PSDs derived for the mean-subtracted simulated light curves using various methods, as denoted in the plots. The left column (panels a-c) in the figure corresponds to the simulated lightcurve with a global monotonic trend, while the right column (panels d-f) corresponds the case of the simulated lightcurve without any global monotonic trend.}       
\label{fig:A1}
\end{figure}

For completeness, we note that the power spectral density (PSD) of an evenly sampled light curve $f(t_i)$ is defined as the rms-normalized squared modulus of the DFT,
\begin{equation}
P(\nu_k) = \frac{2 \, T}{\mu^2 N^2}  \, \left \{ \Bigg[ \sum_{i=1}^{N} f(t_i) \, \cos(2\pi\nu_k t_i)  \Bigg]^2  + \Bigg[ \sum_{i=1}^{N} f(t_i) \, \sin(2\pi\nu_k t_i)  \Bigg]^2 \right \} \, ,
\end{equation}
computed for evenly spaced frequencies $\nu_{k} = k/T$ with $k=1, ..., N/2$, where $T = N (t_k-t_1)/(N-1)$ and $\mu$ is the mean flux. The LS periodogram \citep{S82}, on the other hand, computes the power spectrum as
\begin{equation}
P_f(2 \pi \nu_k )  =  \frac{1}{2}\left \{ \frac{ \Bigg\{ \sum_{i=1}^{N} f(t_i)  \cos[2\pi\nu_k (t_i-\omega)]  \Bigg\}^2} {\sum_{i=1}^{N} f(t_i)  \cos^2[2\pi\nu_k (t_i-\omega)]} 
+  \frac { \Bigg\{ \sum_{i=1}^{N} f(t_i)  \sin[2\pi\nu_k (t_i-\omega)]  \Bigg\}^2} {\sum_{i=1}^{N} f(t_i)  \sin^2[2\pi\nu_k (t_i-\omega)]} \right \} \, ,
\label{eq:lsp}
\end{equation}
where $\omega$ is defined by
\begin{equation}
\omega = \arctan \Bigg\{\frac{1}{2 (2 \pi \nu_k) } \, \frac{ \sum_{i=1}^{N} \sin [2 (2 \pi \nu_k) t_i ]} {  \sum_{i=1}^{N} \cos [ 2 (2 \pi \nu_k) t_i ] }\Bigg\} \, ,  
\label{eq:tau}
\end{equation}
so that the data are equivalent to sum of sines and cosines. Finally, the ACF of the analyzed time series is given by
\begin{equation}
r(\tau) = \frac{E \{ [f(t) - \mu] \times [f(t+\tau) -\mu ] \} } {\sigma_f^2}  \, ,
\label{eq:rtau}
\end{equation}
where $ E\{F\}$ is the expectation value of the function $F$, while $\sigma_f$ denotes the standard deviation of the flux values around the mean $\mu$, and $\tau$ is the `lag' with which the data points are separated \citep{EK88}. The Fourier transform of this ACF gives the spectral power \citep{D75},
\begin{equation}
P(\nu_k) = 2  \sum_{i=1}^{N} r(\tau) \times \cos (2 \pi \nu_k \tau). 
\label{eq:pstau}
\end{equation}

\begin{figure}[t!]
\centering
\includegraphics[width=0.33\textwidth]{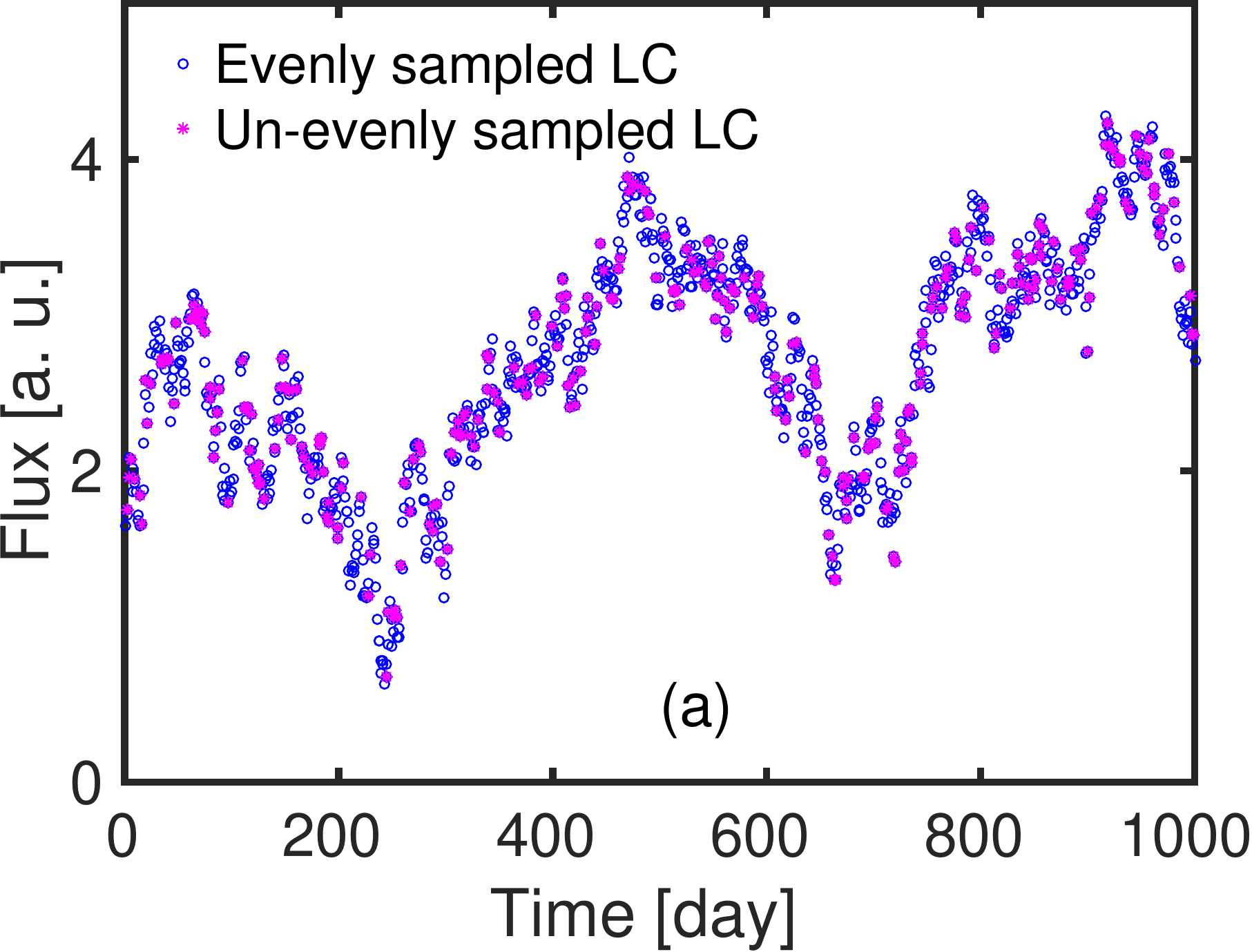}
\includegraphics[width=0.33\textwidth]{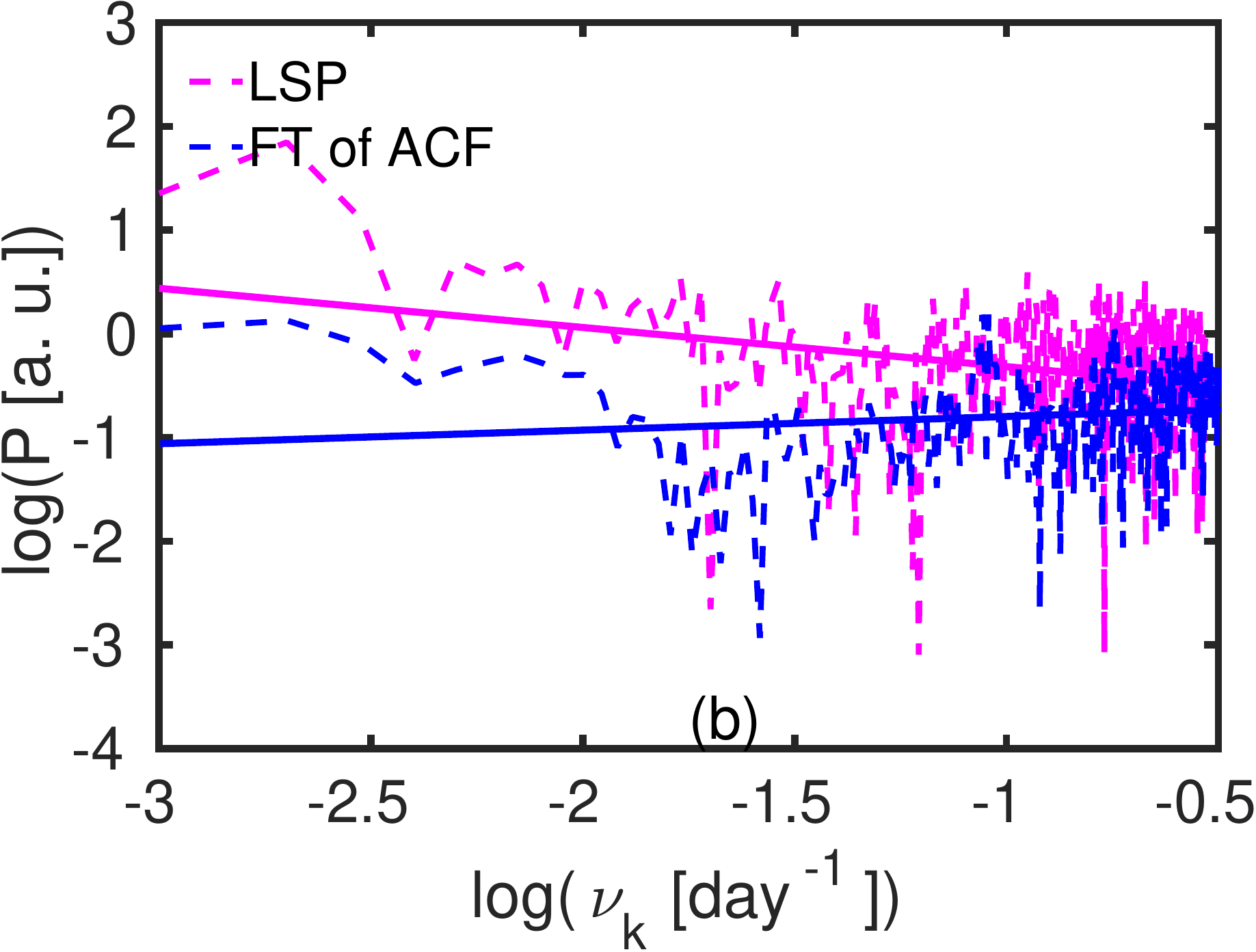}
\includegraphics[width=0.33\textwidth]{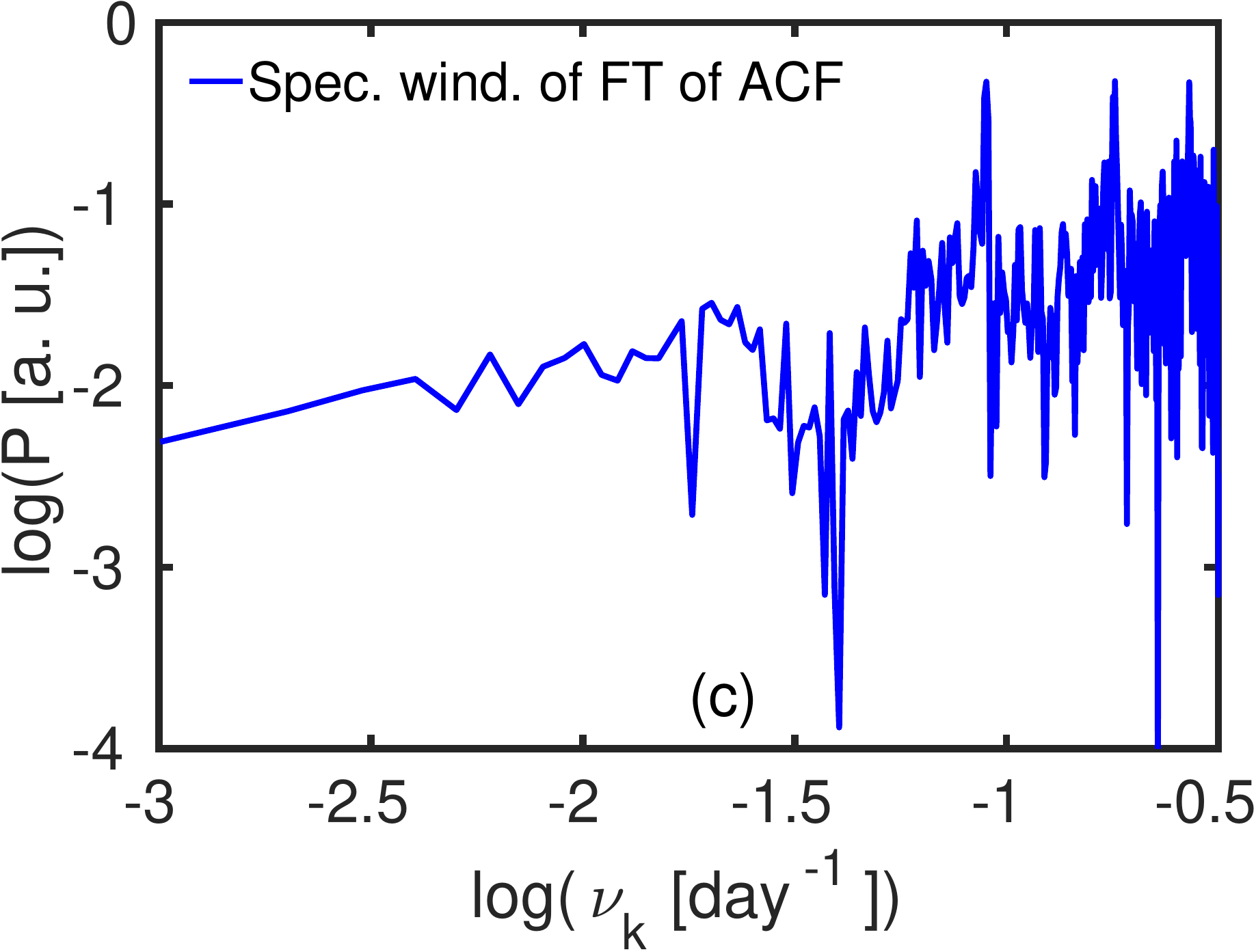}
\caption{{\it Left panel (a):} Simulated red-noise ($\beta=2$) light curve containing 1,000 data points with a sampling period of one day (blue open circles), along with the 30\% of the time series selected at random times to mimic un-evenly sampled dataset (magenta stars). {\it Middle panel (b):} The corresponding power spectra derived using the LS method ($\beta=$ 0.13; magenta dashed curve), and the Fourier transform of the ACF ($\beta= -$0.37; blue dashed curve). {\it Right panel (c):} Spectral window of power spectrum estimated from Fourier transform of the ACF.}
\label{fig:A2}
\end{figure}

At first, we consider a light curve with a global monotonic trend, in order to demonstrate the effect of a power leakage from the side-lobes of the window function. The original simulated light curve, i.e. the light curve before subtracting the mean, is shown in Figure\, \ref{fig:A1}(a); in the panel (b) of the figure, we present the PSDs derived using the LSP method, and the DFT with rectangular, Bartlett, and Hanning window functions for this original light curve.  In  Figure\, \ref{fig:A1}(c), the analogous PSDs derived for the mean-subtracted light curve are given, all in arbitrary units. As shown, the general shape of the PSD is recovered well in the derived peridograms, with slopes $\beta=1.83-1.94$ estimated by fitting straight lines in the log-log space. Here the difference with the true PSD, $\beta = 2$, is exactly due to the red noise leakage effect. In addition, one may see that in the periodograms derived for the original light curve (i.e., the light curve without mean subtraction), ripples arise in the low-frequency segments of the power spectra derived using the DFT methods, depending on the particular shapes of the window functions applied. In principle, this kind of a distortion of the power spectrum should be present only around the 0'th frequency in the power spectrum (as is the case with rectangular window). We found out, however, that with the usage of Bartlett or Hanning windows, lower frequencies (other than the 0'th frequency) are affected as well. These ripples/distortions are minimized in the periodogram derived for the mean-subtracted light curve using the DFT with the Hanning window: as discussed in detail in \citet{2014MNRAS.445..437M}, the Hanning window has the lowest sidelobe level and higher fall-off rate as compared to rectangular and Bartlett windows.

We repeated the analysis for a light curve without any global monotonic trend. The results of the simulations are presented in Figure\, \ref{fig:A1}(d-f). As shown, in this case the red noise leakage effect is less pronounced (although still present), but the low-frequency ripples still arise in the power spectra calculated using the DFT-Bartlett and  DFT-Hanning methods for the original light curve. Again, such distortions are minimized in the case of the DFT-Hanning method applied to the mean-subtracted light curve.

\begin{figure}[t!]
\centering
\includegraphics[width=0.4\textwidth]{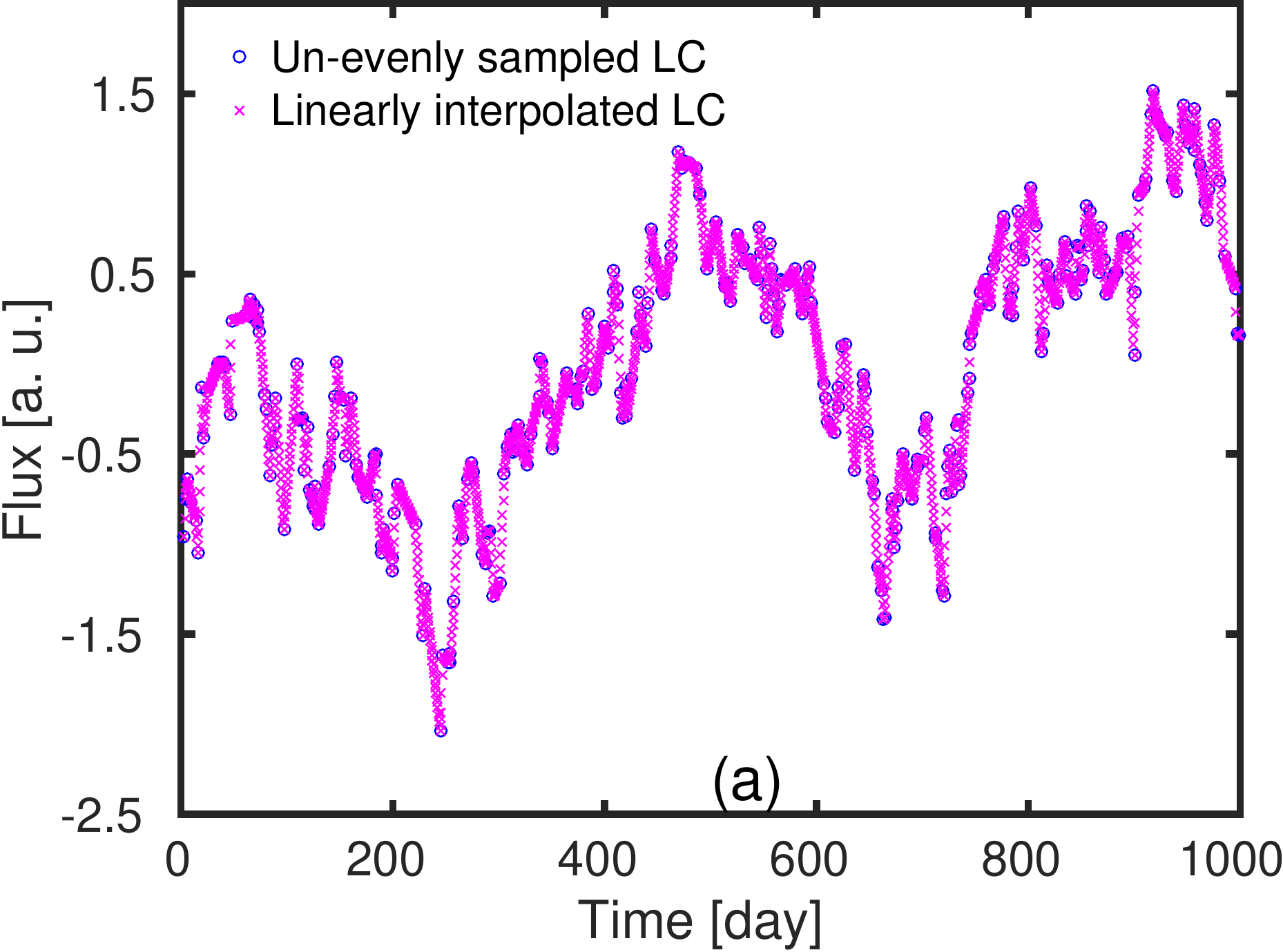}
\includegraphics[width=0.4\textwidth]{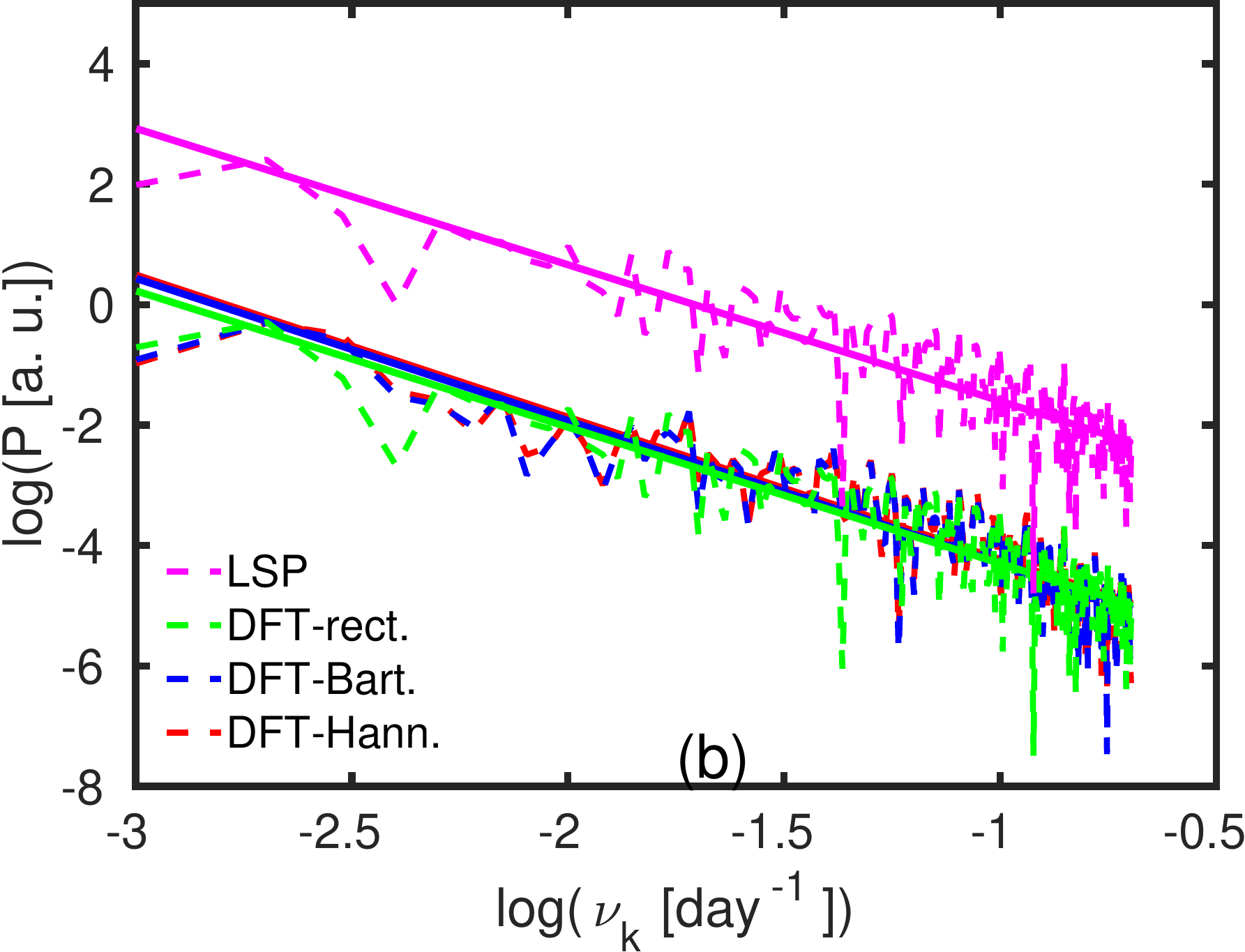}
\caption{{\it Left panel (a):} Unevenly sampled light curve consisting of 300 points (blue open circles), along with the linearly interpolated light curve with a sampling interval of 0.5 day (magenta crosses). {\it Right panel (b):} The corresponding power spectra derived for the interpolated light curve using the LS periodogram, DFT-rectangular, DFT-Bartlett, and  DFT-Hanning methods (fitted slopes $\beta=$ 2.2, 2.2, 2.1 and 2.1, respectively).}
\label{fig:A3}
\end{figure}

Next, we demonstrate that the LS periodogram and Fourier transform of the ACF do not reproduce correctly steep red noise-type power spectra in the case of unevenly sampled data. Generally speaking, in the case of the LS method that is because of a small number of degrees of freedom (DOF) available to characterize the variability at the highest frequencies. On the other hand, in the periodograms derived using Fourier transform of the ACF, one cannot deconvolve the effect of the spectral window, which becomes more and more prominent for less and less evenly spaced data. In order to demonstrate this, we simulated the light curve with $\beta = 2$, containing 1,000 data points with a sampling period of one day; we then subtract the mean and keep 30\% of the data selected at random times to mimic an unevenly sampled dataset; the simulated light curves are presented in the Figure\,\ref{fig:A2}(a). Figure\,\ref{fig:A2}(b) shows the corresponding power spectra derived using the LS method and the Fourier transform of the ACF; the Figure\,\ref{fig:A2}(c) gives the response of the spectral window for the Fourier transform of the ACF. As shown, the derived PSDs (slopes $\beta=$ 0.13 and $-$0.37 for the LS periodogram and the Fourier transform of the ACF, respectively), are very different from the true PSD ($\beta = 2$). The flattening of the derived power spectrum at high variability frequencies in the case of the LS periodogram is due to the aforementioned DOF problem, which introduces  artificial power in the high frequency range of the spectrum. In the case of the Fourier Transform of the ACF, which is free of such a ``missing data points'' problem, the flattening is instead due to a substantial amount of variability power provided by the spectral window. Note in this context that the spectral window does always introduce some extra power in the peridograms derived based on the Fourier transform, but in case of evenly sampled data there is no such extra power at frequencies other than the zeroth Fourier frequency. For unevenly sampled data this extra power becomes, however, significant when compared to the true variability power,  particularly in the highest frequency range; as a result the derived power spectra are artificially flattened. Fourier transforming of the ACF instead of the initial light curve does not provide any solution in this respect, since the window function cannot be deconvolved from the derived power spectrum. 

Finally, we argue that the only method enabling one to produce a robust derivation of the power spectrum for unevenly sampled data, must rely on a linear interpolation of the light curve. In order to do so, we start from the unevenly sampled time series shown in the left panel of Figure\, \ref{fig:A2}, which is then interpolated on a regular grid with a sampling interval of 0.5 day. The Figure\,\ref{fig:A3}(a) shows this original unevenly sampled time series along with the linearly interpolated time series; the corresponding power spectra derived using the LS method and the DFT method with three different window functions are presented in the  Figure\,\ref{fig:A3}(b). The PSD slopes are estimated via linear fitting in the log-log space down to the Nyquist frequency of the original time series ($\sim 6$ days). As shown, all the methods considered reproduce now the true PSD reasonably well. However, keeping in mind the fact that for the time series with a global monotonic trend the DFT-Hanning method provides the best results (in particular against the red noise leakage effect), we conclude that this method is the most robust one.

\vspace{0.5cm}

\twocolumngrid

\bibliographystyle{apj}

\end{document}